\documentclass[epj]{svjour}
\usepackage{graphics}
\usepackage{caption}
\usepackage{subcaption}
\RequirePackage[T1]{fontenc}
\usepackage{float}
\RequirePackage{graphicx}
\RequirePackage{mathptmx}      
\RequirePackage{flushend}
\RequirePackage[numbers,sort&compress]{natbib}
\RequirePackage[colorlinks,citecolor=blue,urlcolor=blue,linkcolor=blue]{hyperref}

\usepackage[numbers]{natbib}
\usepackage{rotating}

\usepackage{graphicx}
\usepackage{braket}
\usepackage{multirow}
\usepackage{amsmath}
\begin{document}
\title{Investigation of Mass and Decay Characteristics of the Light-Strange Tetraquark}
\author{Chetan Lodha\inst{1},
	 \and  Ajay Kumar Rai\inst{2},}                     

%
\institute{Department of Physics, Sardar Vallabhbhai National Institute of Technology, Surat, Gujarat-395007, India, \\ \email{iamchetanlodha@gmail.com }\\ \email{raiajayk@gmail.com}}

\mail{iamchetanlodha@gmail.com}
\date{Received: date / Revised version: date}
%
\abstract{
	Motivated by recent developments in tetraquark studies, we investigate the mass spectra and decay properties of light-strange tetraquarks ($sq\bar{s}\bar{q},ss\bar{q}\bar{q}$) in diquark-antidiquark formalism. By considering different internal quark structures and internal color structures, mass spectra are generated in semi-relativistic and non-relativistic frameworks. For decay widths, the annihilation model and spectator model have been incorporated. Several resonances have been explored as potential candidates for these tetraquarks. Concurrently, mass spectra and several decay channels of kaons ($s\bar{q},q\bar{s}$) are also investigated. This study is carried out in the hope of helping improve the understanding of tetraquarks in the light-light sector. 
\PACS{
      {PACS-key}{discribing text of that key}   \and
      {PACS-key}{discribing text of that key}
     } 
} 

\maketitle
\section{Introduction}

The discovery of the first potential tetraquark candidate in 2003 marks the beginning of a plethora of experimental evidence that compelled physicists worldwide to peer beyond the conventional hadronic structure introduced by Gell-Mann and Zweig in the 1960s \cite{Gell-Mann:1964ewy,Belle:2003nnu}. These unconventional or exotic hadrons were theoretically postulated in the 1980s as a result of the success of the quark model \cite{Jaffe:1976yi,Jaffe:1976ig,Lichtenberg:1978sj,Gignoux:1987cn}. Numerous theoretical models have since been employed to probe these exotic hadrons, including Lattice QCD \cite{Wagner:2013vaa}, QCD Sum Rule \cite{Chen:2016jxd}, potential models, and phenomenological models \cite{Ghalenovi:2020zen,Noh:2023zoq,Mutuk:2023yev,Ortega:2022efc}. These exotic hadrons have been explained by these models as hybrid mesons, tetraquarks, pentaquarks, hadronic molecules, and other structures \cite{Esposito:2016noz, Ali:2017jda, Olsen:2017bmm,Lebed:2016hpi,ParticleDataGroup:2022pth}.

Four-quark resonances have been interpreted as either compact tetraquarks or meson molecules (also referred to as molecular tetraquarks). Numerous studies have proposed that the possible structures of these multiquark hadrons can be understood as singlet states \cite{Chen:2022asf}. Most tetraquark candidates identified so far include at least one charm quark. Notable discoveries include \( Z(4430) \) by Belle \cite{Belle:2007hrb}, \( Y(4660) \) by Belle, \( Y(4140) \) by Fermilab \cite{Mahajan:2009pj}, \( X(5568) \) by the \( D\emptyset \) experiment \cite{D0:2016mwd}, \( Z_{c}(3900) \) by BESIII \cite{BESIII:2013ris}, and \( X(6900) \) by LHCb \cite{LHCb:2020bwg}.

In the light-light tetraquark sector, experimental evidence remains scarce. Research on all-strange and all-light tetraquarks is particularly limited, with only a few studies addressing these configurations. Significant theoretical exploration is still required to better understand the properties and structures of all-strange and all-light tetraquarks.

Recent years have seen significant experimental evidence of strange mesons at facilities such as Belle \cite{Belle:2010xyn}, LHCb \cite{LHCb:2016nsl,LHCb:2021uow}, BaBar \cite{BaBar:2008inr}, and BESIII \cite{BESIII:2019dme}. Strange mesons and their excitations are abundantly observed in \( e^{+}e^{-} \) collisions, hadroproduction, and \(\tau\)-decay. Recent \( e^{+}e^{-} \) collisions at BaBar and BESIII have provided data for the 2.0–2.2 GeV mass range \cite{BESIII:2020kpr,BaBar:2019kds}. Upcoming facilities such as \(\bar{P}\)ANDA, LHCb, Belle II, BESIII, and J-PARC aim to improve measurements of strange mesons through in-depth studies of \(\tau\)-decay \cite{PANDA:2016scz,PANDA:2016fbp,Singh:2016hoh,Singh:2019jug,PANDA:2018zjt,PANDA:2023ljx,PANDA:2021ozp,Aoki:2021cqa}. J-PARC has also planned the construction of a new kaon beamline. These experimental advancements will enhance our understanding of the unconfirmed strange mesons listed in the PDG. Multiple theoretical models have been developed over the years to explain hadron structures \cite{Chaturvedi:2022pmn,Oudichhya:2023lva,Kher:2017mky,Kher:2017wsq,Rai:2008sc,Kher:2018wtv,Purohit:2022mwu,Menapara:2023rur,Menapara:2022ksj}. The current study calculates the mass spectra and \( J^{PC} \) values for the S, P, D, F, and G waves of kaons.

Compact tetraquarks are often described as bound states of diquarks and anti-diquarks. In the present work, the fitting parameters of mesons are obtained and later employed to derive diquarks and tetraquarks. The mass spectra and decay properties of heavy tetraquarks and heavy-light tetraquarks have been extensively studied in non-relativistic formalism. Semi-relativistic formalism, obtained by incorporating relativistic corrections to kinetic energy, has also successfully determined the mass spectra of heavy-light and heavy-heavy mesons. The heavy tetraquark system has been investigated using the bag model, lattice QCD, QCD sum rules, and potential phenomenological models. Similarly, the mass spectra of other exotic states, such as hadronic molecules, pentaquarks, and hybrid mesons, have been studied in detail in non-relativistic formalism \cite{Rai:2006bt}.

Numerous strange closed resonances in the mass range of 1.0–3.5 GeV have been observed in experimental facilities, exhibiting characteristics consistent with four-quark states. Studies suggest that resonances such as \(\eta'\), \( K^{*}(700) \), \( f_{0}(980) \), \( a_{0}(980) \), \( f_{0}(1370) \), and \( f_{0}(1500) \) are better described as four-quark states than conventional two-quark states.

References \cite{Kim:2023bac,Kim:2024adb} propose a tetraquark mixing model that considers light and heavy nonets as tetraquarks formed through the mixing of two tetraquarks. The nature of the \( K^{*}_{0}(700) \) resonance as a tetraquark candidate has been explored in \cite{ALICE:2023eyl} using \( \pi^{\pm}K_{s}^{0} \) correlations. An extensive study of the \( qq\bar{q}\bar{q} \) structure is presented in \cite{Santopinto:2006my}, predicting a tetraquark nonet comprising \( f_{0}(600) \), \( \kappa(800) \), \( f_{0}(980) \), and \( a_{0}(980) \).

Using QCD Sum Rules, \cite{Wang:2006gj} investigates \( X(1576) \) as a tetraquark state in the P-wave configuration with scalar diquarks \([us]\), \([ds]\), \([\bar{u}\bar{s}]\), and \([\bar{d}\bar{s}]\). Similarly, \cite{Xin:2022qnv} employs QCD Sum Rules for an in-depth study of fully light vector tetraquark states, explicitly focusing on P-wave configurations, including all-strange and doubly strange light tetraquarks. 

References \cite{Agaev:2020zad,Agaev:2019coa} explore the vector resonance \( Y(2175) \) as a tetraquark candidate with \( su\bar{s}\bar{u} \) content in a diquark-antidiquark framework using the QCD light-cone sum rule. The present study calculates the mass spectra of S- and P-wave tetraquarks with \( sq\bar{s}\bar{q} \) and \( ss\bar{q}\bar{q} \) structures. Additionally, the decay properties of these tetraquarks are analyzed, along with their possible decay channels.

The present paper is organized as follows: Section II presents the theoretical framework for determining the mass spectra of mesons, diquarks, and tetraquarks, following a brief introduction in Section I. Section III discusses the mass spectroscopy of diquarks, anti-diquarks, tetraquarks, and kaon mesons. The decay properties of kaons and strange tetraquarks are explored in Section IV. Regge trajectories of kaons and tetraquarks are examined in Section V. The findings and discussion are presented in Section VI, with conclusions provided in Section VII.

\section{Theoretical Framework}

The current work is motivated by investigations \cite{Lodha:2024bwn,Lodha:2024qby,Lodha:2023gpp,Tiwari:2021iqu,Tiwari:2021tmz}, where potentials that account for constituent quark interactions for hadrons are inspired by phenomenology. Non-relativistic and semi-relativistic modeling have been used because of the tetraquark's quark composition in the current work. The binding energy of each distinct state is calculated using the modified time-independent radial Schrödinger equation. As described in \cite{Lucha:1995zv}, the center of mass frame is used to incorporate a two-body problem into the central potential model. In order to isolate the radial and angular terms of the time-independent Schrödinger, spherical harmonics have been employed. Using spherical harmonics, the angular and radial terms of the time-independent Schrödinger wave function can be isolated. For mesons and tetraquarks, the fundamental two-body semi-relativistic and non-relativistic Hamiltonian in the center of mass frame, where the motion of constituent particles inside the bound state is relativistic, is given by \cite{Devlani:2011zz,Radford:2009bs}, 

\begin{equation}
	H_{SR} = \sum_{i=1}^2 \sqrt{p^{2}_{i}+M^{2}_{i}} + V^{(0)} (r) + V_{SD}(r),
\end{equation}

\begin{equation}
	H_{NR} = \sum_{i=1}^2 (M_{i} + \frac{p_{i}^{2}}{2M_{i}}) + V^{(0)} (r) + V_{SD}(r).
\end{equation}

where $M_{i}$, $p_{i}$ and V (r) are the constituent mass, relative momentum of the bound state system, and interaction potential, respectively. The kinetic energy term for the non-relativistic Hamiltonian is expanded only up to $\mathcal{O}(p^{2})$. On the other hand, the kinetic energy term of the semi-relativistic Hamiltonian is expanded up to $\mathcal{O}(p^{10})$ as
\begin{equation} 
	\begin{split}
		K.E. = & \sum_{i=1}^2 \frac{p^{2}}{2}\bigr(\frac{1}{M_{i}}\bigr) - \frac{p^{4}}{8}\bigr(\frac{1}{M_{i}^{3}}\bigr) + \frac{p^{6}}{16}\bigr(\frac{1}{M_{i}^{5}}\bigr) \\ 
		&- \frac{5p^{8}}{128}\bigr(\frac{1}{M_{i}^{7}}\bigr) + \frac{7p^{10}}{256}\bigr(\frac{1}{M_{i}^{9}}\bigr)
	\end{split}
\end{equation}

In the expansion up to $\mathcal{O}(p^{10})$, the terms $\mathcal{O}(p^{4})$ and $\mathcal{O}(p^{8})$ exhibit negative values, while $\mathcal{O}(p^{6})$ and $\mathcal{O}(p^{10})$ display positive values. This alternating polarity among the terms suggests a potential cancellation effect, where the positive and negative contributions may partially negate each other. This phenomenon highlights the intricate interplay within the expansion, indicating that certain terms might counterbalance the effects of others, leading to a more refined and accurate description of the system.
Notably, the $\mathcal{O}(p^{10})$ term plays a critical role in shaping the mass spectra of light-strange tetraquarks. Its significance becomes evident when considering the delicate balance between the positive and negative contributions, which collectively influence the overall behavior of these particles. The presence of this higher-order term suggests that the expansion's convergence and the resulting mass spectra are particularly sensitive to the inclusion of higher-order contributions. The nuanced dynamics revealed through this balance underscore the importance of carefully considering all relevant terms in the expansion in accurately capturing the mass distribution of strange tetraquarks.
The investigation of the hadron spectrum has been studied by various inter-quark interaction potentials \cite{Eichten:1979ms,Quigg:1979vr,Godfrey:1985xj}. Heavy quarkonia spectroscopy has been extensively studied in the zeroth-order Cornell-like potential. This potential comprises a Coulombic potential term and a linear potential term. The Coulombic potential term is a result of the Lorentz vector exchange employed in the form of one-gluon exchange, whereas the linear potential term arises due to the Lorentz scalar exchange employed in the form of quark confinement. The zeroth-order Cornell-like potential is given by 

\begin{equation}
	V^{(0)}_{C+L}(r) = \frac{k_{s}\alpha_{s}}{r} + br + V_{0},
\end{equation}

where the QCD running coupling constant, color factor, string tension, and constant are given by $\alpha_{s}$, $k_{s}$, $b$, and $V_{0}$ respectively. The states with color configurations of singlet, triplet-antitriplet, and sextet have color factors of $-\frac{4}{3}, -\frac{2}{3}$, and $\frac{1}{3}$, respectively. Inspired by \cite{Koma:2006si}, relativistic correction to masses has been incorporated into the central potential. Hence, the central potential has the form,

\begin{equation}
	V^{(0)}(r) = V^{0}_{C+L}(r) + V^{1}(r) \biggl(\frac{1}{m_{1}} + \frac{1}{m_{2}} \biggl)
\end{equation}

where the constituent masses of constituent particles in the bound state are given by $m_{1}$ and $m_{2}$. As the non-perturbative form of the relativistic mass correction term is still not studied well enough, the leading order of the perturbation theory is employed \cite{Koma:2006si,Brambilla:2000gk} and is given as

\begin{equation}
	V^{1}(r) = - \frac{C_{F}C_{A}}{4} \frac{\alpha_{s}^{2}}{r^{2}},
\end{equation}

where $C_{F}$ and $C_{A}$ are the Casimir charges of the fundamental and the adjoint representation with values $\frac{4}{3}$ and $3$, respectively. 

The incorporation of spin-dependent terms has been done perturbatively for the model, which provides a better understanding of the splitting between the orbital and radial excitations of various states \cite{Lucha:1991vn}.

\subsection{Spin-dependent Interactions}

The three spin-dependent interactions employed in the present study are inspired by the Breit-Fermi Hamiltonian for one-gluon exchange \cite{Lucha:1991vn,Voloshin:2007dx}. By adding the matrix components of interactions as energy corrections using the first-order perturbation theory, the spin-dependent interactions are determined. The tensor $V_{T}$ interaction potential and spin-orbit interaction potential $V_{LS}$ describe the fine structure of the states, while the hyperfine splitting of the states is described by the spin-spin interaction potential $V_{SS}$.

\begin{equation}
	V_{SD}(r) = V_{T}(r) + V_{LS}(r) + V_{SS}(r).
\end{equation}

In terms of the static potential $V(r)$, the spin-interaction potentials are employed. The tensor interaction potential is defined as:
\begin{equation}
	V_{T} =\; C_{T} \; \biggl(-\frac{1}{3}(S_{1}\cdotp S_{2}) + \frac{(S_{1} \cdotp r) ({S_{2}\cdotp r)}}{r^{2}}\biggl)
\end{equation}
where
\begin{equation}
	C_{T} = - \frac{k_{s}\alpha_{s}}{4} \frac{12\pi}{M_{\mathcal{D}}M_{\bar{\mathcal{D}}}}\frac{1}{r^{3}}
\end{equation}

where $M_{\mathcal{D}}$ and $M_{\bar{\mathcal{D}}}$ represent the masses of quark and antiquark in the case of mesons.  Similarly, in the case of tetraquarks, they represent the masses of diquarks and antidiquarks, respectively. $(S_{1}\cdotp S_{2})$ can be determined by the solution of diagonal matrix elements of spin-$\frac{1}{2}$ and spin-1 particles, as illustrated in ref \cite{Debastiani:2017msn,Lundhammar:2020xvw}. Simplifying, the tensor interaction potential equation is as follows:

\begin{equation}
	S_{12}= 4 \biggl[-(S_{1}\cdotp S_{2}) + 3 (S_{1}\cdotp r)(S_{2}\cdotp r)\biggl]
\end{equation}
The evaluation of \textit{$S_{12}$} is done using Pauli matrices and spherical harmonics with their corresponding eigenvalues. The values of $\braket{S_{12}}$ are hence given as \cite{Tiwari:2021iqu,Tiwari:2021tmz},
\begin{equation}
	\begin{split}
		&  \; -\frac{2l}{2l+3},\text{ for J = l + 1} \\
		\braket{S_{12}}_{\frac{1}{2} \otimes \frac{1}{2} \rightarrow S=1, l \neq 0}= &  \; -\frac{2l+2}{2l-1},\text{ for J = l - 1} \\
		& \; +2, \text{ for J = l}
	\end{split}
\end{equation}

$\braket{S_{12}}$ yields non-zero values for excited states in mesonic states, generalizing $\braket{S_{12}}$ = $-4, +2$ and $\frac{2}{5},$ for J = 0, 1, and 2, respectively, but always vanishes for l = 0 and S = 0. In the case of tetraquarks, spin-1 diquarks are involved, as illustrated in ref \cite{Bethe:1957ncq}.

Employing the same formula as used for mesons, the tensor interaction potential of tetraquarks is obtained, except that the wavefunction obtained here will be of spin-1 (anti) diquark \cite{Debastiani:2017msn,Radford:2009bs,Patel:2016otd}.
\begin{equation}
	\begin{split}
		\braket{S_{d-\bar{d}}}  & =  \; 12  \biggl(-\frac{(S_{d}\cdotp S_{\bar{d}})}{3} + \frac{(S_{d}\cdotp r) {(S_{\bar{d}}\cdotp r)}}{r^{2}}\biggl) \\
		&  \; = S_{14} + S_{13} + S_{24} + S_{23} \\
	\end{split}
\end{equation}

where the total spin of the diquark and anti-diquark are given by $S_{d}$ and $S_{\bar{d}}$, respectively. While obtaining the mass of the tetraquarks by solving the two-body problem, the interaction between the two quarks in the given diquark and the interaction between the two anti-quarks in the anti-diquarks is identical.
The spin-orbit interaction potential is defined as \cite{Lucha:1991vn}:
\begin{equation}
	V_{LS} =\; C_{LS}(L\cdotp S)
\end{equation}
where
\begin{equation}
	C_{LS} =\;  -  \frac{b}{2M_{\mathcal{D}}M_{\bar{\mathcal{D}}}}\frac{1}{r} -  \frac{k_{s}\alpha_{s}}{2}\frac{3\pi}{M_{\mathcal{D}}M_{\bar{\mathcal{D}}}}\frac{1}{r^{2}}
\end{equation}
The first term in $C_{LS}$ is known as Thomas precession. This term arises due to the assumption that the confining interaction originates from a Lorentz scalar structure \cite{Thomas:1926dy,Delll}. This term is proportional to the scalar contribution. 

By incorporating the spin-spin interactions, a suitable approximation can be obtained for mesons and tetraquarks. A $\sigma$ parameter has been introduced as a replacement for the Dirac Delta function. The spin-spin interaction potential is defined as \cite{Lucha:1991vn}:

\begin{equation}
	V_{SS} =\; C_{SS}(S_{1}\cdotp S_{2})
\end{equation}
where
\begin{equation}
	C_{SS} = - \frac{k_{s}\alpha_{s}}{3} \frac{8\pi}{M_{\mathcal{D}}M_{\bar{\mathcal{D}}}} \frac{\sigma}{\sqrt{\pi}}^{3} exp^{-\sigma^{2}r^{2}}
\end{equation}

\begin{figure*}[t]
	\centering
	\begin{subfigure}{0.49\textwidth}
		\includegraphics[width=0.98\linewidth, height=0.29\textheight]{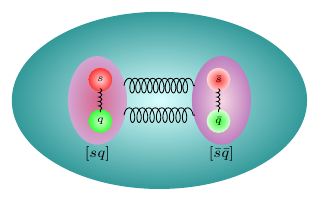}
		\caption{$T_{sq\bar{s}\bar{q}}$}
		\label{fig:tikz---ssqq}
	\end{subfigure}
	\begin{subfigure}{0.49\textwidth}
		\includegraphics[width=0.98\linewidth, height=0.29\textheight]{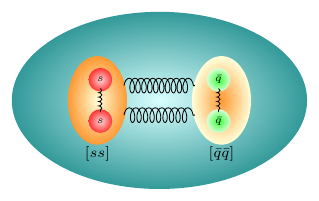}
		\caption{$T_{ss\bar{q}\bar{q}}$}
		\label{fig:tikz---sqsq}
	\end{subfigure}
	\caption{Pictorial representation of tetraquark $T_{sq\bar{s}\bar{q}}$ and $T_{ss\bar{q}\bar{q}}$}
\end{figure*}

For tetraquarks, only the spin-spin interaction is relevant, while the spin-orbit and the tensor interaction are both identically zero considering the fact that the diquarks and anti-diquarks are considered only for the S-wave state. The functional form for spin-$\frac{1}{2}$ particles is based only on the general angular momentum elementary theory \cite{Cohen}. The tensor operator generalized within this approximation can be regarded as the sum of four tensor interactions between four quark-antiquark pairs, which is elucidated in ref \cite{Deba}.
$J^{PC}$ values of different tetraquark states can be obtained by using the relations $P_{T} = (-1)^{L_{T}}$ and $C_{T} = (-1)^{L_{T}+S_{T}}$ where $S_{T}$ and $L_{T}$ are total spin and total angular momentum, respectively \cite{Lucha:1995zv}. Coupling $S_{T}$ and $L_{T}$ results in total angular momentum $J_{T}$, which is used to obtain the mass spectra of radial and orbital excitations. Figure \ref{fig:tikz---sqsq} and Figure \ref{fig:tikz---ssqq} depict the pictorial representations of the tetraquarks \( T_{sq\bar{s}\bar{q}} \) and \( T_{ss\bar{q}\bar{q}} \), respectively.

\section{Spectroscopy}

\subsection{Meson Spectra}

As inspired by previous studies, the mass spectra of kaons ($q\bar{s}$ and $s\bar{q}$) and pions ($q\bar{q}$) are calculated first, and fitting parameters for diquarks and tetraquarks are obtained from these spectra. The fitting parameters $(M_{s},M_{q},\alpha_{s},b,\sigma)$ are calculated in the present work with values $M_{s}=0.515$ $GeV$, $M_{q}=0.37$ $GeV$, $\alpha_{s}=0.9325$, $b=0.125$ $GeV^{2}$ and $\sigma=0.6025$ $GeV$ for kaons ($s\bar{q}$ and $q\bar{s}$). The fitting parameters for pions ($q\bar{q}$) $(M_{q},\alpha_{s},b,\sigma)$ in the present work have values $M_{q}=0.37$ $GeV$, $\alpha_{s}=0.95$, $b=0.1325$ $GeV^{2}$ and $\sigma=0.83$ $GeV$ for pions ($q\bar{q}$). Only colorless quark combinations $\ket{q\bar{q}}$ are permitted to form a color singlet state under SU(3) color symmetry \cite{Deba}. Kaons ($\ket{q\bar{s}}$ and $\ket{s\bar{q}}$) and pions ($\ket{q\bar{q}}$) exhibit $\bar{\textbf{3}}\otimes{\textbf{3}} = \textbf{1} \oplus \textbf{8} $ representation carrying a color factor $k_{s} = -\frac{4}{3} $ \cite{Debastiani:2017msn}. The masses of each Kaon state $(q\bar{s})$ and $s \bar{q})$, the Strangeonium state $[s\bar{s}]$ and the pion $[q\bar{q}]$ states are given by,
\begin{equation}
	\begin{split}
		M_{(s\bar{q})} = &M_{s} + M_{\bar{q}} + E_{(s\bar{q})} + \braket{V^{1}(r)},\\
		M_{(s\bar{s})} = & M_{s} + M_{\bar{s}} + E_{(s\bar{s})} + \braket{V^{1}(r)},\\
		M_{(q\bar{q})} = & M_{q} + M_{\bar{q}} + E_{(q\bar{q})} + \braket{V^{1}(r)}.
	\end{split}
\end{equation} 
where $M_{s}$, $M_{\bar{s}}$, $M_{q}$ and $M_{\bar{q}}$ are constituent masses of strange quark, strange anti-quark, up/down quark, and up/down anti-quark, respectively. Similarly, $E_{(s\bar{q})}$, $E_{(s\bar{s})}$ and $E_{(q\bar{q})}$ are the binding energies of kaons, strangeonium, and pions, respectively. The final mass derived from the above calculation includes contributions from relativistic mass corrections and spin-dependent terms. A decent agreement is seen between the mass spectra generated in the present study and the experimental data from the most recent updated Particle Data Group \cite{ParticleDataGroup:2022pth}. Tables \ref{SWaveMesonmass} and \ref{GWaveMesonmass} include the computed mass for kaons as well as comparisons between several theoretical models and PDG.

\begin{center}
	\begin{table*}
		\caption{Kaon mass spectra and comparison with PDG \cite{ParticleDataGroup:2022pth} and various theoretical models for S, P and D wave in MeV}
		\label{SWaveMesonmass}
		\begin{tabular*}{\textwidth}{@{\extracolsep{\fill}}lrrrrrrrrrrl@{}}
			\hline

			State & \multicolumn{1}{c}{$J^{PC}$} & \multicolumn{1}{c}{Meson} & \multicolumn{1}{c}{Mass$_{NR}$} & \multicolumn{1}{c}{Mass$_{SR}$} & \multicolumn{1}{c}{\cite{ParticleDataGroup:2022pth}} & \cite{Oudichhya:2023lva}   &\cite{Godfrey:1985xj}   &\cite{Ishida:1986vn} &\cite{Vijande:2004he} &\cite{Ebert:2009ub} &\\
			\hline
			&  &  &  &  &&&&&&&\\
			$1 ^{1}S_{0}$ & {$0^{-}$} & &497 & 496 &497.61$\pm$0.013&497.61&462&494&496&482&\\
			$2 ^{1}S_{0}$ &{$0^{-}$} & &1399 &1398 &&1482.40&1454&1360&1472&1538&\\
			$3 ^{1}S_{0}$ &{$0^{-}$} &K(1830) &2011 &1966 &1874$\pm43^{+59}_{-119}$&2036.52&2065&1910&1899&2065&\\
			$4 ^{1}S_{0}$ &{$0^{-}$} & &2515 &2454 &&2469.27&&2210&&&\\
			&&&&&&&&&&&\\
			$1 ^{3}S_{1}$ &$1^{-}$ & &891 &899 &891.67$\pm$0.26&891.67&903&898&910&897&\\
			$2 ^{3}S_{1}$ &$1^{-}$ & &1606 &1588 &&1675.00&1579&1670&1620&1675&\\
			$3 ^{3}S_{1}$ &$1^{-}$ & &2161 &2094&&2194.57&1950&2190&&2156&\\
			$4 ^{3}S_{1}$ &$1^{-}$ & &2637 &2529&&2612.79&&2490&&&\\
			
			&  &  &  & &&& &&&&\\
			
			\hline
			&  &  &  &  &&&&&&&\\
			
			$1 ^{1}P_{1}$&$1^{+}$ &$K_{1}(1400)$ &1383	&1391&$1403\pm7$&1286.81&1352&1370&1372&1294&\\
			$2 ^{1}P_{1}$ &$1^{+}$ & &1958	&1917&&1757.00&1897&1980&1841&1757&\\
			$3 ^{1}P_{1}$ &$1^{+}$ & & 2454 &2370&&2125.61&2164&2440&&2164&\\
			$4 ^{1}P_{1}$ &$1^{+}$ & &2896 &2808 &&2439.14&&&&&\\
			$5 ^{1}P_{1}$ &$1^{+}$ & &&&&&&&&&\\
			&&&&&&&&&&&\\
			$1 ^{3}P_{0}$ &$0^{+}$  & &1094	&1092 &&1362.00&1234&1060&1213&1362&\\
			$2 ^{3}P_{0}$ &$0^{+}$ &K(1630) &1686	&1645 &$1630\pm7$&1791.00&1890&1730&1768&1791&\\
			$3 ^{3}P_{0}$ &$0^{+}$ & &2188	&2103 &&2135.49&2160&2220&&2160&\\
			$4 ^{3}P_{0}$ &$0^{+}$ & &2634	&2527 &&2431.66&&&&&\\
			&&&&&&&&&&&\\
			$1 ^{3}P_{1}$ &$1^{+}$  & &1418	&1438 &&1403.00&1366&1340&1394&1412&\\
			$2 ^{3}P_{1}$ &$1^{+}$ & &1998	&1969&&1893.00&1928&1960&1850&1893&\\
			$3 ^{3}P_{1}$ &$1^{+}$ & &2497	&2417&&2280.11&2200&2430&&2200&\\
			$4 ^{3}P_{1}$ &$1^{+}$ & &2939	&2835 &&2610.39&&&&&\\
			&&&&&&&&&&&\\
			$1 ^{3}P_{2}$ &$2^{+}$  &$K_{2}^{*}(1430)$ &1454	&1491&$1432.4\pm1.3$&1419.20&1428&1390&1450&1424&\\
			$2 ^{3}P_{2}$ &$2^{+}$ &$K_{2}^{*}(1980)$ &2045	&2028 &$1994^{+60}_{-50}$&1994.00&1938&2020&&1896&\\
			$3 ^{3}P_{2}$ &$2^{+}$ & &2549	&2479 &&2436.79&2206&2490&&2206&\\
			$4 ^{3}P_{2}$ &$2^{+}$ & &2996	&2900 &&2810.66&&&&&\\
			
			\hline
			&  &  &  &  &&&&&&&\\
			
			$1 ^{1}D_{2}$&$2^{-}$  &$K_{2}(1770)$ &1742	&1727&&1750.46&1791&1760&1747&1709&\\
			$2 ^{1}D_{2}$ &$2^{-}$ &$K_{2}(2250)$ &2249	&2184&$2247\pm17$&2066.00&2238&2290&&2066&\\
			$3 ^{1}D_{2}$ &$2^{-}$ & &2705 &2602&&2339.35&&&&&\\
			$4 ^{1}D_{2}$ &$2^{-}$ &&3121	&3035&&2583.95&&&&&\\
			&&&&&&&&&&&\\
			$1 ^{3}D_{1}$ &$1^{-}$ &$K^{*}(1680)$&1708	&1695 &$1718\pm18$&1700.77&1776&1680&1698&1699&\\
			$2 ^{3}D_{1}$ &$1^{-}$ &&2218	&2157 &&2063.00&2251&2210&&2063&\\
			$3 ^{3}D_{1}$ &$1^{-}$ &&2676	&2577 &&2370.51&&&&&\\
			$4 ^{3}D_{1}$ &$1^{-}$ &$K(3100)$&3094	&3004 &3100&2642.47&&&&&\\
			&&&&&&&&&&&\\
			$1 ^{3}D_{2}$ &$2^{-}$  &&1740	&1737 &&1756.84&1804&1740&1741&1824&\\
			$2 ^{3}D_{2}$ &$2^{-}$ &$K_{2}(2250)$& 2257	&2204&$2247\pm17$&2163.00&2254&2280&&2163&\\
			$3 ^{3}D_{2}$ &$2^{-}$ && 2720	&2627 &&2504.13&&&&&\\
			$4 ^{3}D_{2}$ &$2^{-}$ && 3142	&3057&&2804.06&&&&&\\
			&&&&&&&&&&&\\
			$1 ^{3}D_{3}$ &$3^{-}$  &&1710	&1712 &&1798.11&1794&1760&1766&1789&\\
			$2 ^{3}D_{3}$ &$3^{-}$  &&2236	&2210 &&2182.00&2237&2300&&2182&\\
			$3 ^{3}D_{3}$ &$3^{-}$  &&2705	&2652 &&2507.80&&&&&\\
			$4 ^{3}D_{3}$ &$3^{-}$  &&3130	&3052 &&2795.88&&&&&\\
			\hline
			
		\end{tabular*}
	\end{table*}
\end{center}

\begin{center}
	\begin{table*}
		\caption{Kaon mass spectra and comparison with PDG \cite{ParticleDataGroup:2022pth} and various theoretical models for F and G wave  in MeV}
		\label{GWaveMesonmass}
		\begin{tabular*}{\textwidth}{@{\extracolsep{\fill}}lrrrrrrrrrrl@{}}
			\hline

			State & \multicolumn{1}{c}{$J^{P}$} & \multicolumn{1}{c}{Meson} & \multicolumn{1}{c}{Mass$_{NR}$} & \multicolumn{1}{c}{Mass$_{SR}$} & \multicolumn{1}{c}{\cite{ParticleDataGroup:2022pth}} & \cite{Oudichhya:2023lva}   &\cite{Godfrey:1985xj}   &\cite{Ishida:1986vn} &\cite{Vijande:2004he} &\cite{Ebert:2009ub} &\\
			\hline
			&  &  &  &  &&&&&&&\\
			
			$1 ^{1}F_{3}$&$3^{+}$&&2043&2010&&2114.79&2131&2047&&2009&\\
			$2 ^{1}F_{3}$&$3^{+}$&&2510&2429&&&&2340&&&\\
			$3 ^{1}F_{3}$&$3^{+}$&&2939&2839&&&&2507&&&\\
			
			&&&&&&&&&&&\\
			$1 ^{3}F_{2}$&$2^{+}$&$K_{2}^{*}(1980)$&2003&1980&$1994^{+60}_{-50}$&1982.46&2151&2095&1968&1964&\\
			$2 ^{3}F_{2}$&$2^{+}$&&2476&2404&&&&2364&&&\\
			$3 ^{3}F_{2}$&$2^{+}$&&2910&2818&&&&2519&&&\\
			&&&&&&&&&&&\\
			$1 ^{3}F_{3}$&$3^{+}$&&1994&1980&&2050.5&2143&2132&&2080&\\
			$2 ^{3}F_{3}$&$3^{+}$&&2474&2408&&&&2387&&&\\
			$3 ^{3}F_{3}$&$3^{+}$&&2912&2825&&&&2529&&&\\
			&&&&&&&&&&&\\
			$1 ^{3}F_{4}$&$4^{+}$&&1958&2014&&2110.04&2108&2080&&2096&\\
			$2 ^{3}F_{4}$&$4^{+}$&&2444&2470&&&&2359&&&\\
			$3 ^{3}F_{4}$&$4^{+}$&&2886&2884&&&&2517&&&\\
			\hline
			&  &  &  &  &&&&&&&\\
			
			$1 ^{1}G_{4}$&$4^{-}$&&2323&2264&&2424.99&2422&2270&&2255&\\
			$2 ^{1}G_{4}$&$4^{-}$&&2759&2663&&&&2476&&&\\
			
			&&&&&&&&&&&\\
			$1 ^{3}G_{3}$&$3^{-}$&&2259&2214&&2228.84&2458&2316&&2207&\\
			$2 ^{3}G_{3}$&$3^{-}$&&2704&2620&&&&2499&&&\\
			
			&&&&&&&&&&&\\
			$1 ^{3}G_{4}$&$4^{-}$&&2240&2203&&2307.08&2433&2337&&2285&\\
			$2 ^{3}G_{4}$&$4^{-}$&&2689&2612&&&&2510&&&\\
			
			&&&&&&&&&&&\\
			$1 ^{3}G_{5}$&$5^{-}$&$K_{5}^{*}(2380)$&2212&2346&$2382\pm19\pm14$&2381.46&2388&2291&&2356&\\
			$2 ^{3}G_{5}$&$5^{-}$&&2665&2770&&&&2488&&&\\

			\hline
			
		\end{tabular*}
	\end{table*}
\end{center}		

\subsection{Diquarks}

A bound state of two quarks or two anti-quarks interacting with each other through gluonic exchange is generally perceived as a diquark $(\mathcal{D})$ and an anti-diquark ($\mathcal{\bar{D}}$), respectively \cite{Ebert:2007rn}. Diquarks and anti-diquarks interact, resulting in composite systems rather than point-like objects. As a result of Pauli's exclusion principle, the ground-state wavefunction is of an antisymmetric nature. The diquark's nature can be purely scalar, axial-vector or vector; however, diquarks composed of identical quark flavors can only have spin S = 1. Scalar diquarks are dubbed "good diquarks," while axial-vector diquarks are "bad diquarks," which is discussed in \cite{Esposito:2016noz} in greater detail.
The method used to calculate the masses of diquarks and anti-diquarks is the same as for kaons and pions. The fundamental representation for a diquark in QCD color symmetry is represented by $\textbf{3}\otimes\textbf{3}=\bar{\textbf{3}}\oplus \textbf{6}$ fundamental $(\textbf{3})$ representation  \cite{Debastiani:2017msn}. Likewise, an anti-diquark in the $\bar{\textbf{3}}$ representation is given by $\bar{\textbf{3}}\otimes\bar{\textbf{3}}=\textbf{3}\oplus\bar{\textbf{6}}$. A tetraquark is considered a four-body problem since it is composed of two quarks and two anti-quarks. This four-body problem is reduced to a two-body problem by employing the diquark-antidiquark approximation \cite{Fredriksson:1981mh}. $\textbf{1}\otimes{\textbf{1}}$ state and the $\textbf{8}\otimes\bar{\textbf{8}}$ state are formed due to the $\bar{\textbf{3}}\otimes{\textbf{3}}$ color coupling. As the color factor $k_{s}$ in QCD color symmetry for the triplet-antitriplet state is $-\frac{2}{3}$, the short distance part, $\frac{1}{r}$ of the interaction has an attractive nature \cite{Deba}. The masses of diquarks and anti-diquarks studied in the present work are given by,    
\begin{equation}
	\begin{split}
		M_{(ij)} = &M_{i} + M_{j} + E_{(ij)} + \braket{V^{1}(r)}\\
		M_{(\bar{i}\bar{j})} = &M_{\bar{i}} + M_{\bar{j}} + E_{\bar{i}\bar{j}} + \braket{V^{1}(r)}\\	
	\end{split}
\end{equation} 
where $M_{i}$ and $M_{j}$ are masses of constituent quarks in the diquark, while $M_{\bar{i}}$ and $M_{\bar{j}}$ are masses of constituent anti-quarks in the anti-diquark. The mass of all the diquarks calculated in the present work is tabulated in table \ref{diquark}.

\begin{table*}	
	\centering
	\caption{Mass spectra of various diquarks/ anti-diquarks. A comparison with various theoretical models is made. All units are in MeV.}
	\label{diquark}
	\begin{tabular}{ccccccccccc}
		
		\hline
		\multirow{2}{*}{Diquark}&\multicolumn{2}{c}{ Mass$_{NR}$} &\multicolumn{2}{c}{Mass$_{SR}$} & \multirow{2}{*}{\cite{Maris:2002yu}} &\multirow{2}{*}{\cite{Faustov:2021hjs}}  &\multirow{2}{*}{\cite{Chen:2023ngj}} &\multirow{2}{*}{\cite{Ferretti:2019zyh}}
		&\multirow{2}{*}{\cite{Yin:2021uom}} & \multirow{2}{*}{\cite{Hess:1998sd}} \\ 
		&Triplet &Sextet &Triplet &Sextet&&&&&&\\
		\hline
		{$qq$} &989&1130  &942 &1126  &1020  &909     &970 &840&1060&730 \\
		
		{$sq$} &1106 &1194 &1090 &1209 &1306  &1069  &1110 &992&1160&\\
		
		{$ss$} & 1236 &1350 &1215 &1215  &1444  &1203  &1240&1136&1260&1210\\
		\hline
	\end{tabular}	
\end{table*}

\subsection{Tetraquark Spectra}
A color singlet tetraquark can be materialized by two different diquark-antidiquark combinations: (i) a color anti-triplet diquark and a color triplet anti-diquark, or (ii) a color sextet diquark and a color anti-sextet anti-diquark. Each of these two combinations, held together by color force, forms a tetraquark in a singlet configuration. A tetraquark in the singlet state for $\bar{\textbf{3}}-\textbf{3}$ and $\textbf{6}-\bar{\textbf{6}}$ has a color factor $k_{s}$ = $- \frac{4}{3}$ and $-\frac{10}{3}$, respectively\cite{Deba,Debastiani:2017msn}. By combining spin-1 diquark and anti-diquark, a singlet tetraquark can be formed. This singlet teraquark has the representation $\ket{QQ|^{3}\otimes|\bar{Q}\bar{Q}|^{\bar{3}}} = \textbf{1}\oplus\textbf{8}$. A tetraquark with two strange quarks and two up/down quarks can have more than one internal structure based on the quark flavor of the diquarks or anti-diquarks involved. A $[ss]$ diquark fused with $[qq]$ anti-diquark will result in $T_{ss\bar{q}\bar{q}}$, while a $[sq]$ diquark fused with $[sq]$ anti-diquark will result in $T_{sq\bar{s}\bar{q}}$. The $T_{sq\bar{s}\bar{q}}$ tetraquark resembles a quarkonia-like structure, while the $T_{ss\bar{q}\bar{q}}$ resembles a kaonic structure. Hence, the mass spectra for $T_{sq\bar{s}\bar{q}}$ and $T_{ss\bar{q}\bar{q}}$ have the formulation,
\begin{equation}
	\begin{split}
		M_{(SQ\bar{S}\bar{Q})} = & M_{SQ} + M_{\bar{S}\bar{Q}} + E_{(SQ\bar{S}\bar{Q})} + \braket{V^{1}(r)}\\		
		M_{(SS\bar{Q}\bar{Q})} = & M_{SS} + M_{\bar{Q}\bar{Q}} + E_{(SS\bar{Q}\bar{Q})} + \braket{V^{1}(r)}.		
	\end{split}
\end{equation} 
The Cornell-like potential $V^{0}_{C+L}$, the relativistic term $\braket{V^{1}(r)}$, and the spin-dependent contributions contribute to the mass spectra of the calculated tetraquark states. All the spin-dependent contributions are calculated individually. By coupling total spin $S_{T}$ with orbital angular momentum $L_{T}$, $S_{T}\otimes L_{T}$, the color singlet state of the tetraquark is obtained.
\begin{equation}
	\ket{T} = \ket{S_{d},S_{\bar{d}},S_{T},L_{T}}_{J_{T}},
\end{equation}
where $S_{d}$ and $S_{\bar{d}}$ are the spins of diquark and anti-diquark, respectively. In the case of mesons, diquarks, and anti-diquarks, only two spin combinations were possible. However, in the case of a tetraquark, three spin combinations are possible while using spin-1 diquarks and anti-diquarks.

\begin{equation} 
	   \begin{split}        \ket{0^{++}}_{T}=\ket{S_{d}=1,S_{\bar{d}}=1,S_{T}=0,L_{T}=0}_{J_{T=0}};\\        \ket{1^{+-}}_{T}=\ket{S_{d}=1,S_{\bar{d}}=1,S_{T}=1,L_{T}=0}_{J_{T=1}};\\        \ket{2^{++}}_{T}=\ket{S_{d}=1,S_{\bar{d}}=1,S_{T}=2,L_{T}=0}_{J_{T=2}};    \end{split}
\end{equation}
Utilizing the one-gluon exchange between a quark from a diquark and an anti-quark from an anti-diquark${\textbf{6}}\otimes\bar{\textbf{6}} $ and$\bar{\textbf{3}}\otimes{\textbf{3}} $ states can be mixed. However, for this mixing to occur, a four-body problem approach for tetraquarks must be considered. Since the current work employs the diquark-antidiquark approximation to treat the tetraquark as a two-body problem, a mixed-state tetraquark cannot be formed. On the other hand, the ${\textbf{6}}\otimes\bar{\textbf{6}} $ configuration is found to be highly suppressed when compared to the $\bar{\textbf{3}}\otimes{\textbf{3}}$ configuration \cite{Park:2013fda}. Due to the repulsive nature of the sextet diquark, the $\bar{\textbf{3}}\otimes{\textbf{3}}$ component in the composite wave function completely dominates the ${\textbf{6}}\otimes\bar{\textbf{6}}$ component. A four-body approach suggests that $\bar{\textbf{3}}\otimes{\textbf{3}}$ and ${\textbf{6}}\otimes\bar{\textbf{6}}$ states behave as two individual states even though they can mix  \cite{Debastiani:2017msn}. The mass spectra of $\bar{\textbf{3}}\otimes{\textbf{3}}$ and ${\textbf{6}}\otimes\bar{\textbf{6}}$ states for S wave and P wave for $ss\bar{q}\bar{q}$ and $sq\bar{s}\bar{q}$ are done in tables \ref{Swavetriplet} and \ref{Swavesextet}, respectively. The various two-meson thresholds possible for $ss\bar{q}\bar{q}$ and $sq\bar{s}\bar{q}$ have been tabulated in table \ref{twomesonthreshold}. The parameter sensitivity of the current theoretical model for mass spectra has been discussed in great detail in our previous work \cite{Lodha:2024qby}.  

\begin{table}
	\centering
	\caption{Mass spectra of Tetraquarks $T_{ss\bar{q}\bar{q}}$ and $T_{sq\bar{s}\bar{q}}$ with $\bar{3}\otimes{3}$ diquark-antidiquark configuration. A comparison with various theoretical models as well as two-meson thresholds is made. All units are in MeV.}
	\label{Swavetriplet}
	\begin{tabular}{cccccc}
		
		\hline
		\multirow{2}{*}{State} &\multirow{2}{*}{$J^{PC}$} &\multicolumn{2}{c}{Semi-Relativistic Mass} &\multicolumn{2}{c}{Non-Relativistic Mass} \\
		&& $sq\bar{s}\bar{q}$ & $ss\bar{q}\bar{q}$ & $sq\bar{s}\bar{q}$ & $ss\bar{q}\bar{q}$  \\
		\hline
		&&&&&\\
		$1 ^{1}S_{0}$ &	$0^{++}$ &1563	&1552 &1760&1767\\
		$2 ^{1}S_{0}$&$0^{++}$ 	&2591 &2573 &2716&2723\\
		$3 ^{1}S_{0}$&$0^{++}$ 	&3065 &3049 &3205&3211\\
		&&&&&\\
		$1 ^{3}S_{1}$ &$1^{+-}$	&1734 &1724 &1918&1926\\
		$2 ^{3}S_{1}$&$1^{+-}$	&2633 &2618 &2760&2768\\
		$3 ^{3}S_{1}$&$1^{+-}$	&3093 &3078&3232&3241\\
		&&&&&\\
		$1 ^{5}S_{2}$ &	{$2^{++}$}	&2077 &2071 &2236 &2243\\
		$2 ^{5}S_{2}$&{$2^{++}$} 	&2717 &2705 &2848 &2855\\
		$3 ^{5}S_{2}$&{$2^{++}$} 	&3147 &3135 &3288 &3297\\
		\hline
		
		&&&&&\\
		$1 ^{1}P_{1}$&	$1^{--}$&2506 &2489 &2632 &2637\\
		$2 ^{1}P_{1}$&$1^{--}$ 	&2968 &2951 &3099 &3105\\
		$3 ^{1}P_{1}$&$1^{--}$	&3333 &3319 &3481 &3489\\
		&&&&&\\
		$1 ^{3}P_{0}$ &	$0^{-+}$	&2147 &2133 &2305 &2315\\
		$2 ^{3}P_{0}$&$0^{-+}$ 	&2666  &2651  &2823  &2832\\
		$3 ^{3}P_{0}$&$0^{-+}$	&3055  &3041  &3226 &3234 \\
		&&&&&\\
		$1 ^{3}P_{1}$ &	$1^{-+}$	&2515  &2499 &2636 &2644 \\
		$2 ^{3}P_{1}$&$1^{-+}$	&2970  &2955  &3101 &3109 \\
		$3 ^{3}P_{1}$&$1^{-+}$	&3332 &3318  &3481 &3489 \\
		&&&&&\\
		$1 ^{3}P_{2}$ &	$2^{-+}$	&2634 &2615  &2740 &2747 \\
		$2 ^{3}P_{2}$&$2^{-+}$	&3071  &3055 &3191 &3200\\
		$3 ^{3}P_{2}$&$2^{-+}$	&3426 &3412 &3566 &3574 \\
		&&&&&\\
		$1 ^{5}P_{1}$ &	$1^{--}$ &2153  &2139  &2304 &2314 \\
		$2 ^{5}P_{1}$&$1^{--}$	&2665 &2651 &2820 &2829\\
		$3 ^{5}P_{1}$&$1^{--}$	&3048   &3035 &3219 &3228\\
		&&&&&\\
		$1 ^{5}P_{2}$ &$2^{--}$	&2615  &2598  &2719 &2727\\
		$2 ^{5}P_{2}$&$2^{--}$	&3047  &3033 &3169 &3177 \\
		$3 ^{5}P_{2}$&$2^{--}$	&3398 &3385  &3540 &3548 \\
		&&&&&\\
		$1 ^{5}P_{3}$&	$3^{--}$	&2793  &2774  &2875 &2881\\
		$2 ^{5}P_{3}$&$3^{--}$	&3198  &3183 &3304 &3312 \\
		$3 ^{5}P_{3}$&$3^{--}$	&3538  &3525  &3667 &3675 \\
		\hline
	\end{tabular}
\end{table}
\begin{table}
	\centering
	\caption{Mass spectra of Tetraquarks $T_{ss\bar{q}\bar{q}}$ and $T_{sq\bar{s}\bar{q}}$ with $\bar{3}\otimes{3}$ diquark-antidiquark configuration in \textbf{6}-$\bar{\textbf{6}}$. A comparison with various theoretical models as well as two-meson thresholds is made. All units are in MeV.}
	\label{Swavesextet}
	\begin{tabular}{cccccc}
		
		\hline
		\multirow{2}{*}{State} &\multirow{2}{*}{$J^{PC}$} &\multicolumn{2}{c}{Semi-Relativistic Mass} &\multicolumn{2}{c}{Non-Relativistic Mass} \\
		&& $sq\bar{s}\bar{q}$ & $ss\bar{q}\bar{q}$ & $sq\bar{s}\bar{q}$ & $ss\bar{q}\bar{q}$  \\
		\hline
		&&&&&\\
		$1 ^{1}S_{0}$ &	$0^{++}$ 	&1234	&1299  &795 &853 \\
		$2 ^{1}S_{0}$&$0^{++}$ 	&2711  &2796  &3019 &3064 \\
		$3 ^{1}S_{0}$&$0^{++}$ 	&3762  &3754  &3971 &4019 \\
		\hline
		
		&&&&&\\
		$1 ^{1}P_{1}$&	$1^{--}$&2581  &2568  &2777 &2826 \\
		$2 ^{1}P_{1}$&$1^{--}$ 	&3586  &3573  &3763&3810\\
		$3 ^{1}P_{1}$&$1^{--}$	&4290  &4283 &4499 &4546 \\
		&&&&&\\
		$1 ^{3}P_{0}$ &	$0^{-+}$	&634 &628  &1040&1107 \\
		$2 ^{3}P_{0}$&$0^{-+}$ 	&2084  &2070  &2383&2451\\
		$3 ^{3}P_{0}$&$0^{-+}$	&2965  &2953  &3264 &3330 \\
		&&&&&\\
		$1 ^{3}P_{1}$ &	$1^{-+}$	&2550 &2537 &2752 &2803 \\
		$2 ^{3}P_{1}$&$1^{-+}$	&3525  &3512 &3715&3763 \\
		$3 ^{3}P_{1}$&$1^{-+}$	&4223  &4215  &4446 &4491 \\
		&&&&&\\
		$1 ^{3}P_{2}$ &	$2^{-+}$	&3229  &3213  &3345 &3390 \\
		$2 ^{3}P_{2}$&$2^{-+}$	&4040  &4027  &4182 &4224 \\
		$3 ^{3}P_{2}$&$2^{-+}$	&4679 &4671  &4865 &4905 \\
		\hline
	\end{tabular}
\end{table}

\begin{table}
	\centering
	\caption{Two meson threshold for different states of tetraquark $ss\bar{q}\bar{q}$ and $sq\bar{s}\bar{q}$}
	\label{twomesonthreshold}
	\begin{tabular}{cccc}
		\hline	
		{State} & {Two-meson Threshold} &{Threshold Mass (Semi-relativistic)}&{ Threshold Mass (Non-relativistic)}\\  
		\hline
		\multirow{2}{*}{$^{1}S_{0}$} & $K_{0}^{\pm}(1S)K^{\mp}_{0}(1S)$ &995& 992  \\
		& $\eta_{s}(1S)$ $\pi(1S)$ &882 & 900 \\
		\hline
		\multirow{3}{*}{$^{3}S_{1}$} & $K_{1}^{*}(1S)K_{0}^{\pm}(1S)$ &1389 &1396\\
		&  $\phi(1S)$ $\pi(1S)$ &1159 &1159   \\
		& $\rho(1S)$ $\eta_{s}(1S)$ &1518 &1533\\
		\hline
		\multirow{2}{*}{$^{5}S_{2}$} & $K_{1}^{*}(1S)K^{*}_{1}(1S)$ &1798& 1782  \\
		& $\rho(1S)$ $\phi(1S)$ &1795 & 1791 \\
		\hline
		\multirow{3}{*}{$^{3}P_{0}$} & $K_{0}(1S)$  $K_{0}(1P)$ &1592 &1588\\
		&  $\eta_{s}(1S)$  $a_{0}(1P)$  &1641&1618  \\
		& $\pi(1S)$  $f_{0}(1P)$ &1393&1386\\
		\hline
		\multirow{3}{*}{$^{3}P_{1}$} & $K_{0}(1S)$  $K_{1}(1P)$ &1915 &1934 \\
		&  $\eta_{s}(1S)$  $a_{1}(1P)$  &2055& 2059 \\
		& $\pi(1S)$  $f_{1}(1P)$&1593 &1597 \\
		
		\hline
		\multirow{3}{*}{$^{3}P_{2}$} & $K_{0}(1S)$  $K_{2}(1P)$&1952 &1987  \\
		&  $\eta_{s}(1S)$  $a_{2}(1P)$  &2108 &2092   \\
		& $\pi(1S)$  $f_{2}(1P)$ &1626&1631 \\
		\hline
		\multirow{3}{*}{$^{5}P_{1}$} & $K_{0}(1S)$  $K_{1}(1P)$&1881&1887 \\
		&  $\eta_{s}(1S)$  $b_{1}(1P)$  &2018 &2020   \\
		& $\pi(1S)$  $h_{1}(1P)$ &1572&1575 \\
		\hline
		\multirow{3}{*}{$^{5}P_{2}$} & $K^{*}_{1}(1S)$  $K_{1}(1P)$ &2310 &2338 \\
		&  $\phi(1S)$  $a_{1}(1P)$   &2332&2317  \\
		& $\rho(1S)$  $f_{1}(1P)$ &2229 &2229 \\
		\hline
		\multirow{3}{*}{$^{5}P_{3}$} & $K^{*}_{1}(1S)$  $K_{2}(1P)$ &2346 &2391\\
		&  $\phi(1S)$  $a_{2}(1P)$   &2384 &2351  \\
		& $\rho(1S)$  $f_{2}(1P)$ &2262&2263  \\
		\hline
	\end{tabular}
\end{table}

	\section{Decay}

	Decay widths of bound states entail important properties about their internal structures. As a four-body problem, the dynamics of the decay mechanism for tetraquarks are seemingly very complex. The present work studies the decay properties of kaons as well as the tetraquark states. For kaons, various decay channels involving leptonic and radiative decays are studied. For tetraquarks, rearrangement phenomena and strong decay are employed, inspired by ref. \cite{Becchi:2020mjz,Becchi:2020uvq,Lodha:2024qby}. Following that, the diquark-antiannihilationillation model is employed, which is an extension to quark-anannihilationillation for mesons \cite{Devlani:2011zz}, for gluonic, leptonic, and photonic decay channels. Lastly, the indirect  decay of tetraquark by leptonic, photonic, and similar decay channels of the hadrons from the rearrangement model is calculated.
		
	\subsection{Annihilation decay}	
	Annihilation decay for tetraquarks treats the diquark and anti-diquarks in a manner similar to the quark-antiquark annihilation for heavy quarkonia. Hence, the same formulation has been extended to the tetraquark model, giving us leptonic, gluonic, and photonic decay. The annihilation decay for $T_{^{3}S_{1}}$ and $T_{^{1}S_{0}} $ tetraquark states in ${\bar{3}\otimes3} $ diquark-antidiquark color configuration and $T_{^{1}S_{0}} $ tetraquark states in ${\bar{3}\otimes3}$ diquark-antidiquark color configuration ${6\otimes\bar{6}}$ has been tabulated in table \ref{annihilationdecay}. The total decay rate for a tetraquark with a diquark-antidiquark color configuration ${\bar{3}\otimes3} $ and ${6\otimes\bar{6}}$ are tabulated in table \ref{totaldecay}.
	
	An essential quantity in the study of hadron decay processes is the square modulus of the wave function at the origin, $|\psi(0)|^2$, which is particularly important for calculating decay widths. In quantum mechanics, the wave function or its derivative at the origin plays a critical role in determining the properties of bound states. Specifically, for quarkonium models, the Schrödinger equation includes a centrifugal term that introduces what is known as a "centrifugal barrier." This barrier limits the calculation of the wave function at the origin to states with orbital angular momentum l = 0 (S-states), where the wave function does not vanish. For excited states with $l \neq 0$, the centrifugal barrier causes the wave function at the origin to vanish, making it impossible to calculate $|\psi(0)|^2$ directly for these states. The relationship can be expressed as \cite{Lucha:1991vn}:
	
	\[
	|\psi(0)|^2 = |Y_{00}(\theta, \phi)R_{nl}(0)|^2 = \frac{|R_{nl}(0)|^2}{4\pi}
	\]
	
		Here, $|R_{nl}(0)|^2$ represents the square modulus of the radial wave function at the origin, which can be obtained through numerical calculations.

	The $^{1}S_{0}$ $sq\bar{s}\bar{q}$ tetraquark annihilates into 2 gluons, while the $^{3}S_{1}$ $sq\bar{s}\bar{q}$ tetraquark annihilates into 3 gluons. The gluonic decay width, including the first-order radiative correction, is given by \cite{Segovia:2016xqb,Kwong:1987ak,Kwong:1988ae,Belanger:1987cg,Lodha:2024erj}:

	\begin{equation}
		\Gamma_{ ^{1}S_{0} \; sq\bar{s}\bar{q}\rightarrow g g} = \frac{2 \alpha_{s}^{2} |R_{nl}(0)|^{2}}{3m_{sq}^{2}}\biggr(1+\frac{4.8 \alpha_{s}}{\pi}\biggr),
	\end{equation}
	
	\begin{equation}
		\Gamma_{^{3}S_{1} \; sq\bar{s}\bar{q}\rightarrow ggg} = \frac{10(\pi^{2}-9) \alpha_{s}^{3} |R_{nl}(0)|^{2}}{81\pi m_{sq}^{2}},
	\end{equation}
	where $((1+\frac{4.8 \alpha_{s}}{\pi})$ is the QCD correction factors for the digluonic decay channels.

	The $^{3}S_{1}$ state of tetraquark $sq\bar{s}\bar{q}$ can annihilate into lepton pairs with decay width given by \cite{Wang:2019tqf}:
	\begin{equation}
		\Gamma_{sq\bar{s}\bar{q}\rightarrow e^{+}e^{-}} = \frac{4\pi \alpha^{2} e_{Q}^{2}f_{sq\bar{s}\bar{q}}}{3M_{sq\bar{s}\bar{q}}}\times\biggr(1+2\frac{m_{e}^{2}}{M_{sq\bar{s}\bar{q}}^{2}}\biggr)\sqrt{1-4\frac{m_{e}^{2}}{M_{sq\bar{s}\bar{q}}^{2}}}
	\end{equation}
	\begin{equation}
		\Gamma_{sq\bar{s}\bar{q}\rightarrow \mu^{+}\mu^{-}} = \frac{4\pi \alpha^{2} e_{Q}^{2}f_{sq\bar{s}\bar{q}}}{3M_{sq\bar{s}\bar{q}}}\times\biggr(1+2\frac{m_{\mu}^{2}}{M_{sq\bar{s}\bar{q}}^{2}}\biggr)\sqrt{1-4\frac{m_{\mu}^{2}}{M_{sq\bar{s}\bar{q}}^{2}}}
	\end{equation}
	
where $\alpha$ is the fine structure constant, $e_{Q}$ is the electric charge of the diquark in units of the electron charge, $m_{e}$ is the mass of the electron, $m_{\mu}$ is the mass of the muon, and $M_{sq\bar{s}\bar{q}}$ is the mass of the $sq\bar{s}\bar{q}$ tetraquark. 

\(|R_{nl}(0)|^2\) plays a crucial role in the analysis of pseudoscalar and vector states, as it directly impacts the calculation of decay constants. These constants are essential for understanding meson decay properties and predicting their branching ratios. Studies, such as \cite{Ebert:2006hj}, have explored the decay constants of light-heavy mesons using relativistic treatments, which provide a detailed description of systems influenced by relativistic effects. 

In this work, we estimate the decay constants for the vector state by evaluating the radial wave function at the origin. This approach offers a simplified yet effective method for analyzing tetraquark decay processes and allows for meaningful comparisons with results from more sophisticated relativistic models.
The formula is provided by Van-Royen and Weisskopf \cite{VanRoyen:1967nq, Lodha:2024qby}:
\begin{equation}
	f^{2}_{sq\bar{s}\bar{q}}  = \frac{3|R_{nlV}(0)|^{2}}{\pi M_{nlV}} 	
\end{equation}

where \(|R_{nlV}(0)|^{2}\) and \(M_{nlV}\) are the wave function at the origin and the mass of the vector state, respectively. 

Moreover, these decay constants are critical for theoretical predictions and comparisons with experimental data, aiding in the validation or refinement of tetraquark behavior models. By accurately calculating \(|\psi(0)|^2\) and decay constants, one can gain valuable insights into the strong interaction, the binding energy, and the dynamics of diquark-antidiquark systems. 

The \(^{1}S_{0}\) \(sq\bar{s}\bar{q}\) tetraquark annihilates into two photons, while the \(^{3}S_{1}\) \(sq\bar{s}\bar{q}\) tetraquark annihilates into three photons. The photonic decay width, incorporating first-order radiative corrections, is given by \cite{Segovia:2016xqb, Kwong:1987ak}:

	\begin{equation}
		\Gamma_{sq\bar{s}\bar{q}\rightarrow \gamma \gamma} = \frac{3 \alpha^{2} e_{Q}^{4}|R_{nl}(0)|^{2}}{m_{sq}^{2}},
	\end{equation}
	
	\begin{equation}
		\Gamma_{sq\bar{s}\bar{q}\rightarrow \gamma\gamma \gamma} = \frac{4(\pi^{2}-9) \alpha^{3} e_{Q}^{6}|R_{nl}(0)|^{2}}{3\pi m_{sq}^{2}},
	\end{equation}
	where $m_{sq}$ is the constituent mass of the diquark and $|R_{nl}(0)|^{2}$ is the square modulus of the radial wave function at the origin.

	\begin{table}
		\centering
		\caption{Annihilation decay rates for various tetraquark channels in MeV}
		\label{annihilationdecay}
		\begin{tabular}{cccc}
			\hline	
			{Internal color configuration}	&{Decay Channel} & {Semi-relativistic} & {Non-relativistic}  \\ 
			\hline
			\multirow{7}{*}{${\bar{3}\otimes3}$}&$T_{^{3}S_{1}} \rightarrow e^{+}e^{-}$ & 0.188 & $0.118$\\
			&$T_{^{3}S_{1}} \rightarrow\mu^{+}\mu^{-}$ & $0.188$ & $0.118$ \\
			
			&$T_{^{1}S_{0}} \rightarrow2 \gamma$ & $0.044$ & $0.045$\\
			&$T_{^{3}S_{1}} \rightarrow$ $3\gamma$ & $39.517\times10^{-6}$ & $40.719\times10^{-6}$ \\
			&$T_{^{1}S_{0}} \rightarrow 2g$ & 398.776 & 328.397   \\
			&$T_{^{3}S_{1}} \rightarrow 3g$ & 8.097 & 6.303  \\
			\hline
			\multirow{2}{*}{${6\otimes\bar{6}}$} &$T_{^{1}S_{0}} \rightarrow 2g$ & 1023.99 & 811.348   \\
			&$T_{^{1}S_{0}} \rightarrow 2\gamma$ & $0.116$ & $0.108$   \\
			
			\hline
		\end{tabular}
		
	\end{table}
	\begin{table*}
		\centering
		\caption{Total Decay of $1 ^{1}S_{0}$ $T_{sq\bar{s}\bar{q}}$ and $T_{ss\bar{q}\bar{q}}$ tetraquark in MeV}
		\label{totaldecay}
		\begin{tabular}{ccc}
			\hline
			{Internal color configuration}   & Semi-relativistic & Non-relativistic   \\
			\hline
			$\bar{3} \otimes 3$ & 407.3  &334.9  \\
			$6 \otimes\bar{6}$ &1024.1  &811.5   \\
			\hline
		\end{tabular}
	\end{table*}

	\subsection{Strong Decay}
A tetraquark described using diquark-antidiquark formalism can undergo strong decay through the fall-apart mechanism into the corresponding mesons. The partial decay widths for this process across different channels are given by:
	\begin{equation}
		\Gamma_{T\rightarrow V_{1}V_{2}}= \mathcal{G}_{TV_{1}V_{2}}^{2}  \frac{\lambda(T,V_{1},V_{2})}{8\pi}\biggr(\frac{m_{V_{1}}^{2}m_{V_{2}}^{2}}{m_{T}^{2}}+\frac{2\lambda^{2}(T,V_{1},V_{2})}{3} \biggr)
	\end{equation}

	\begin{equation}
		\Gamma_{T\rightarrow S_{1}S_{2}}= \mathcal{G}_{T S_{1}S_{2}}^{2}  \frac{\lambda(T,S_{1},S_{2})}{8\pi}\biggr( m_{S_{1}}m_{S_{2}} + \lambda^{2}(T,S_{1},S_{2})  \biggr)
	\end{equation} 
	
	\begin{equation}
		\lambda(m_{1},m_{2},m_{3}) = \frac{\sqrt{m_{1}^{4}+m_{2}^{4}+m_{3}^{4}-2(m_{1}^{2}m_{2}^{2}+m_{2}^{2}m_{3}^{2}+m_{1}^{2}m_{3}^{2})}}{2m_{1}}
	\end{equation}

where $V_{1},V_{2},S_{1}$ and $S_{2}$ are decaying vector and scalar mesons, while $\mathcal{G}_{TV_{1}V_{2}}$ and $\mathcal{G}_{TS_{1}S_{2}}$ are the coupling constants for the processes $T\rightarrow V_{1}V_{2}$ and $T\rightarrow S_{1}S_{2}$ respectively.
While comprehensive experimental data on the decay mechanisms of tetraquarks is still limited, several significant theoretical studies have made important strides in addressing this topic. In our previous work, referenced in \cite{Lodha:2024qby}, we explored the strong decay processes for all-strange tetraquarks. Similarly, the potential decay channels for all-strange tetraquarks were investigated and discussed  in  reference \cite{Drenska:2008gr,Jiang:2023atq}. Additionally, the authors in references \cite{Agaev:2023gaq, Agaev:2023wua, Agaev:2023ruu} employed QCD sum rule methods to assess the strong decay characteristics and coupling constants of all-charm and all-bottom tetraquark states. Moreover, ref. \cite{Lu:2019ira} delves into radiative transitions, analyzing the ratios of branching ratios across various decay channels for all-strange tetraquarks, thereby providing valuable theoretical insights into their decay properties. These studies significantly enhance our theoretical understanding of exotic tetraquark systems, shedding light on their decay mechanisms and fundamental properties within the framework of quantum chromodynamics (QCD). The calculated results for various decay channels are summarized in Table \ref{decay}, offering a comprehensive overview of the current theoretical landscape. 
	\begin{center}
		\begin{table}[H]
			\centering
			\caption{Partial decay width for strong decay for various decay channels of all strange tetraquark in semi-relativistic and non-relativistic formalism }
			\label{decay}
			\begin{tabular}{ccccc}
				\hline
				Decay Channel & $\lambda_{SR}(GeV)$  &$\lambda_{NR}(GeV)$  &$\Gamma_{SR}(GeV)$  &$\Gamma_{NR}(GeV)$  \\
				\hline
				$T_{ssqq}(0^{++})_{\bar{\textbf{3}}-\textbf{3}}\rightarrow K_{0} K_{0}$ & 0.5959 &0.7311  &0.02273$\mathcal{G}_{T_{ssqq}K_{0}K_{0}}^{2}$  &0.01425$\mathcal{G}_{T_{ssqq}K_{0}K_{0}}^{2}$  \\
				
				$T_{ssqq}(2^{++})_{\bar{\textbf{3}}-\textbf{3}}\rightarrow K_{1}^{*} K_{1}^{*}$ &0.5139  &0.6811  &0.00671$\mathcal{G}_{T_{ssqq}K_{1}^{*}K_{1}^{*}}^{2}$  &0.01177$\mathcal{G}_{T_{ssqq}K_{1}^{*}K_{1}^{*}}^{2}$  \\
				
				$T_{ssqq}(0^{++})_{\textbf{6}-\bar{\textbf{6}}}\rightarrow K_{0} K_{0}$ & 0.4193 &-  &0.00703$\mathcal{G}_{T_{ssqq}K_{0}K_{0}}^{2}$  &-  \\

				$T_{sqsq}(0^{++})_{\bar{\textbf{3}}-\textbf{3}}\rightarrow K_{0} K_{0}$ &0.6031  &0.7269  &0.01467$\mathcal{G}_{T_{sqsq}K_{0}K_{0}}^{2}$  &0.02239$\mathcal{G}_{T_{sqsq}K_{0}K_{0}}^{2}$  \\
				
				$T_{sqsq}(2^{++})_{\bar{\textbf{3}}-\textbf{3}}\rightarrow K_{1}^{*} K_{1}^{*}$ &0.5199  &0.6753  &0.03712$\mathcal{G}_{T_{4s}\eta\eta}^{2}$  &0.01179$\mathcal{G}_{T_{sqsq}K_{1}^{*}K_{1}^{*}}^{2}$  \\
								
				$T_{sqsq}(0^{++})_{\textbf{6}-\bar{\textbf{6}}}\rightarrow K_{0} K_{0}$ &0.3670  &-  &0.00555$\mathcal{G}_{T_{sqsq}K_{0}K_{0}}^{2}$  &-\\

				$T_{sqsq}(0^{++})_{\bar{\textbf{3}}-\textbf{3}}\rightarrow \pi \eta_{s}$ &0.5937  & 0.7078 &0.01071$\mathcal{G}_{T_{sqsq}\pi\eta_{s}}^{2}$  &0.01708$\mathcal{G}_{T_{sqsq}\pi\eta_{s}}^{2}$  \\
				
				$T_{sqsq}(2^{++})_{\bar{\textbf{3}}-\textbf{3}}\rightarrow \phi \rho$ &0.6957  &0.8010  &0.01120$\mathcal{G}_{T_{sqsq}\phi\rho}^{2}$  &0.01591$\mathcal{G}_{T_{4s}\phi\rho}^{2}$  \\
				
				$T_{sqsq}(0^{++})_{\textbf{6}-\bar{\textbf{6}}}\rightarrow \pi \eta_{s}$ &0.3747  & - &0.00359$\mathcal{G}_{T_{sqsq}\pi\eta_{s}}^{2}$  &-  \\

				\hline
			\end{tabular}	
		\end{table}
	\end{center}

	\subsection{Re-arrangement Decay}
	
	\begin{figure*}[t]
		\centering
		\begin{subfigure}{0.32\textwidth}
			\includegraphics[width=0.98\linewidth, height=0.29\textheight]{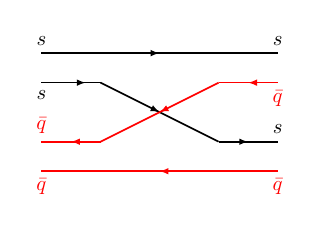}
			\caption{$T_{ss\bar{q}\bar{q}}\rightarrow K + K$}
		\end{subfigure}
		\begin{subfigure}{0.32\textwidth}
			\includegraphics[width=0.98\linewidth, height=0.29\textheight]{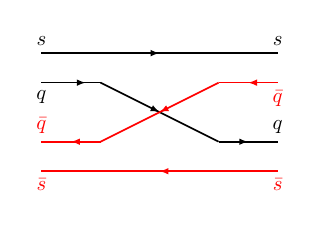}
			\caption{$T_{sq\bar{s}\bar{q}}\rightarrow K + \bar{K}$}
		\end{subfigure}
		\begin{subfigure}{0.32\textwidth}
			\includegraphics[width=0.98\linewidth, height=0.29\textheight]{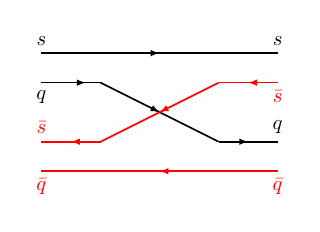}
			\caption{$T_{sq\bar{s}\bar{q}}\rightarrow \pi+\phi$}
		\end{subfigure}
		
		\caption[]{Rearrangement diagram for Tetraquarks $T_{ss\bar{q}\bar{q}}$ and $T_{sq\bar{s}\bar{q}}$}
		\label{reaarange}
	\end{figure*}
	The recoupling of spin wave functions is given by:
	
	\begin{equation}		
		\biggr|{\{{(ss)}^{1}{(\bar{q}\bar{q})}^{1}\}^{0}}\biggr \rangle=-\frac{1}{2} \biggr|\bigr(s\bar{q}\bigr)^{1}\bigr(s\bar{q}\bigr)^{1}\biggr \rangle ^{0} + \frac{\sqrt{3}}{2}\biggr|\bigr(s\bar{q}\bigr)^{0}\bigr(s\bar{q}\bigr)^{0}\biggr \rangle ^{0} 	
		\label{spinwf1}		
	\end{equation}

	\begin{equation}
		\begin{split}		
			\biggr|{\{{(sq)}^{1}{(\bar{s}\bar{q})}^{1}\}^{0}}\biggr \rangle=&\frac{1}{2}\biggr(-\frac{1}{2} \biggr|\bigr(s\bar{q}\bigr)^{1}\bigr(s\bar{q}\bigr)^{1}\biggr \rangle ^{0} + \frac{\sqrt{3}}{2}\biggr|\bigr(s\bar{q}\bigr)^{0}\bigr(s\bar{q}\bigr)^{0}\biggr \rangle ^{0} \biggr)\\ + &\frac{1}{2}\biggr(-\frac{1}{2} \biggr|\bigr(s\bar{s}\bigr)^{1}\bigr(q\bar{q}\bigr)^{1}\biggr \rangle ^{0} + \frac{\sqrt{3}}{2}\biggr|\bigr(s\bar{s}\bigr)^{0}\bigr(q\bar{q}\bigr)^{0}\biggr \rangle ^{0} \biggr)
		\end{split} 	
		\label{spinwf2}		
	\end{equation}
	Similarly, the recoupling of color wave functions is given by:
	\begin{equation}		
		|{{(ss)}_{\bar{\textbf{3}}}{(\bar{q}\bar{q})}_{\textbf{3}}}|\rangle=\sqrt{\frac{1}{3}} |\bigr(s\bar{q}\bigr)_{\textbf{1}}\bigr(s\bar{q}\bigr)_{\textbf{1}}\rangle - \sqrt{\frac{2}{3}}|\bigr(s\bar{q}\bigr)_{\textbf{8}}\bigr(s\bar{q}\bigr)_{\textbf{8}}\rangle 
		\label{colorwf33_1}		
	\end{equation}
	
	\begin{equation}		
		\begin{split}
			|{{(sq)}_{\bar{\textbf{3}}}{(\bar{s}\bar{q})}_{\textbf{3}}}|\rangle= & \frac{1}{2} \biggr( \sqrt{\frac{1}{3}} |\bigr(s\bar{s}\bigr)_{\textbf{1}}\bigr(q\bar{q}\bigr)_{\textbf{1}}\rangle - \sqrt{\frac{2}{3}}|\bigr(s\bar{s}\bigr)_{\textbf{8}}\bigr(q\bar{q}\bigr)_{\textbf{8}}\rangle \biggr)    + \\
			& \frac{1}{2} \biggr( \sqrt{\frac{1}{3}} |\bigr(s\bar{q}\bigr)_{\textbf{1}}\bigr(s\bar{q}\bigr)_{\textbf{1}}\rangle - \sqrt{\frac{2}{3}}|\bigr(s\bar{q}\bigr)_{\textbf{8}}\bigr(s\bar{q}\bigr)_{\textbf{8}}\rangle \biggr)
		\end{split}
		\label{colorwf33_2}		
	\end{equation}

	\begin{equation}		
		|{{(ss)}_{\textbf{6}}}{(\bar{q}\bar{q})}_{\bar{\textbf{6}}}|\rangle=\sqrt{\frac{2}{3}} |\bigr(s\bar{q}\bigr)_{\textbf{1}}\bigr(s\bar{q}\bigr)_{\textbf{1}}\rangle + \sqrt{\frac{1}{3}}|\bigr(s\bar{q}\bigr)_{\textbf{8}}\bigr(s\bar{q}\bigr)_{\textbf{8}}\rangle 
		\label{colorwf66_1}		
	\end{equation}

	\begin{equation}	
		\begin{split}	
			|{{(sq)}_{\textbf{6}}}{(\bar{s}\bar{q})}_{\bar{\textbf{6}}}|\rangle= & \frac{1}{2} \biggr(\sqrt{\frac{2}{3}} |\bigr(s\bar{s}\bigr)_{\textbf{1}}\bigr(q\bar{q}\bigr)_{\textbf{1}}\rangle + \sqrt{\frac{1}{3}}|\bigr(s\bar{s}\bigr)_{\textbf{8}}\bigr(q\bar{q}\bigr)_{\textbf{8}}\rangle \biggr) + \\ & \frac{1}{2} \biggr(\sqrt{\frac{2}{3}} |\bigr(s\bar{q}\bigr)_{\textbf{1}}\bigr(s\bar{q}\bigr)_{\textbf{1}}\rangle + \sqrt{\frac{1}{3}}|\bigr(s\bar{q}\bigr)_{\textbf{8}}\bigr(s\bar{q}\bigr)_{\textbf{8}}\rangle \biggr)
		\end{split}
		\label{colorwf66_2}		
	\end{equation}
	
	Utilizing the Fierz transformation, various quark-antiquark pairs ($s\bar{s}$,$s\bar{q}$,$q\bar{q}$) are brought together \cite{Ali:2019roi}, and by using the spectator pair method, the tetraquark decays into two mesons. The quark bilinears are normalized to unity. The Fierz re-arrangement for various states employing equations for spin wave function recoupling and color wave function recoupling is given by, 
	
	\begin{eqnarray}
		\begin{split}
			&ss\bar{q}\bar{q}(J=0)=\biggr|\bigr(ss\bigr)^{1}_{\bar{\textbf{3}}}\bigr(\bar{q}\bar{q}\bigr)^{1}_{\textbf{3}}\biggr \rangle ^{0}_{1} \\ 
			&=-\frac{1}{2} \biggr(\sqrt{\frac{1}{3}}\biggr|\bigr(s\bar{q}\bigr)^{1}_{\textbf{1}}\bigr(s\bar{q}\bigr)^{1}_{\textbf{1}}\biggr \rangle ^{0}_{\textbf{1}}  -\sqrt{\frac{2}{3}}\biggr|\bigr(s\bar{q}\bigr)^{1}_{\textbf{8}}\bigr(s\bar{q}\bigr)^{1}_{\textbf{8}}\biggr \rangle ^{0}_{\textbf{1}}  \biggr)\\
			&+\frac{\sqrt{3}}{2} \biggr(\sqrt{\frac{1}{3}}\biggr|\bigr(s\bar{q}\bigr)^{0}_{\textbf{1}}\bigr(s\bar{q}\bigr)^{0}_{\textbf{1}}\biggr \rangle ^{0}_{\textbf{1}}  -\sqrt{\frac{2}{3}}\biggr|\bigr(s\bar{q}\bigr)^{0}_{\textbf{8}}\bigr(s\bar{q}\bigr)^{0}_{\textbf{8}}\biggr \rangle ^{0}_{1}  \biggr)
		\end{split}
		\label{eqssqq}
	\end{eqnarray}
	
	\begin{eqnarray}
		\begin{split}
			&sq\bar{s}\bar{q}(J=0)=\biggr|\bigr(sq\bigr)^{1}_{\bar{\textbf{3}}}\bigr(\bar{s}\bar{q}\bigr)^{1}_{\textbf{3}}\biggr \rangle ^{0}_{1} \\ 
			&=-\frac{1}{4} \biggr(\sqrt{\frac{1}{3}}\biggr|\bigr(s\bar{q}\bigr)^{1}_{\textbf{1}}\bigr(s\bar{q}\bigr)^{1}_{\textbf{1}}\biggr \rangle ^{0}_{\textbf{1}}  -\sqrt{\frac{2}{3}}\biggr|\bigr(s\bar{q}\bigr)^{1}_{\textbf{8}}\bigr(s\bar{q}\bigr)^{1}_{\textbf{8}}\biggr \rangle ^{0}_{\textbf{1}}  \biggr)\\
			&+\frac{\sqrt{3}}{4} \biggr(\sqrt{\frac{1}{3}}\biggr|\bigr(s\bar{q}\bigr)^{0}_{\textbf{1}}\bigr(s\bar{q}\bigr)^{0}_{\textbf{1}}\biggr \rangle ^{0}_{\textbf{1}}  -\sqrt{\frac{2}{3}}\biggr|\bigr(s\bar{q}\bigr)^{0}_{\textbf{8}}\bigr(s\bar{q}\bigr)^{0}_{\textbf{8}}\biggr \rangle ^{0}_{1}  \biggr)\\ 
			&-\frac{1}{4} \biggr(\sqrt{\frac{1}{3}}\biggr|\bigr(s\bar{s}\bigr)^{1}_{\textbf{1}}\bigr(q\bar{q}\bigr)^{1}_{\textbf{1}}\biggr \rangle ^{0}_{\textbf{1}}  -\sqrt{\frac{2}{3}}\biggr|\bigr(s\bar{s}\bigr)^{1}_{\textbf{8}}\bigr(q\bar{q}\bigr)^{1}_{\textbf{8}}\biggr \rangle ^{0}_{\textbf{1}}  \biggr)\\
			&+\frac{\sqrt{3}}{4} \biggr(\sqrt{\frac{1}{3}}\biggr|\bigr(s\bar{s}\bigr)^{0}_{\textbf{1}}\bigr(q\bar{q}\bigr)^{0}_{\textbf{1}}\biggr \rangle ^{0}_{\textbf{1}}  -\sqrt{\frac{2}{3}}\biggr|\bigr(s\bar{s}\bigr)^{0}_{\textbf{8}}\bigr(q\bar{q}\bigr)^{0}_{\textbf{8}}\biggr \rangle ^{0}_{1}  \biggr)
		\end{split}
		\label{eqsqsq}
	\end{eqnarray}
	\begin{eqnarray}
		\begin{split}
			&ss\bar{q}\bar{q}(J=0)=\biggr|\bigr(ss\bigr)^{1}_{\textbf{6}}\bigr(\bar{q}\bar{q}\bigr)^{1}_{\bar{\textbf{6}}}\biggr \rangle ^{0}_{\textbf{1}} \\ 
			&=-\frac{1}{2} \biggr(\sqrt{\frac{2}{3}}\biggr|\bigr(s\bar{q}\bigr)^{1}_{\textbf{1}}\bigr(s\bar{q}\bigr)^{1}_{\textbf{1}}\biggr \rangle ^{0}_{\textbf{1}}  +\sqrt{\frac{1}{3}}\biggr|\bigr(s\bar{q}\bigr)^{1}_{\textbf{8}}\bigr(s\bar{q}\bigr)^{1}_{\textbf{8}}\biggr \rangle ^{0}_{\textbf{1}}  \biggr)\\
			&+\frac{\sqrt{3}}{2} \biggr(\sqrt{\frac{2}{3}}\biggr|\bigr(s\bar{q}\bigr)^{0}_{\textbf{1}}\bigr(s\bar{q}\bigr)^{0}_{\textbf{1}}\biggr \rangle ^{0}_{\textbf{1}}  +\sqrt{\frac{1}{3}}\biggr|\bigr(s\bar{q}\bigr)^{0}_{\textbf{8}}\bigr(s\bar{q}\bigr)^{0}_{\textbf{8}}\biggr \rangle ^{0}_{\textbf{1}}  \biggr)
		\end{split}
		\label{eq25}
	\end{eqnarray}
	
	\begin{eqnarray}
		\begin{split}
			&sq\bar{s}\bar{q}(J=0)=\biggr|\bigr(sq\bigr)^{1}_{\textbf{6}}\bigr(\bar{s}\bar{q}\bigr)^{1}_{\bar{\textbf{6}}}\biggr \rangle ^{0}_{\textbf{1}} \\ 
			&=-\frac{1}{4} \biggr(\sqrt{\frac{2}{3}}\biggr|\bigr(s\bar{q}\bigr)^{1}_{\textbf{1}}\bigr(s\bar{q}\bigr)^{1}_{\textbf{1}}\biggr \rangle ^{0}_{\textbf{1}}  +\sqrt{\frac{1}{3}}\biggr|\bigr(s\bar{q}\bigr)^{1}_{\textbf{8}}\bigr(s\bar{q}\bigr)^{1}_{\textbf{8}}\biggr \rangle ^{0}_{\textbf{1}}  \biggr)\\
			&+\frac{\sqrt{3}}{4} \biggr(\sqrt{\frac{2}{3}}\biggr|\bigr(s\bar{q}\bigr)^{0}_{\textbf{1}}\bigr(s\bar{q}\bigr)^{0}_{\textbf{1}}\biggr \rangle ^{0}_{\textbf{1}}  +\sqrt{\frac{1}{3}}\biggr|\bigr(s\bar{q}\bigr)^{0}_{\textbf{8}}\bigr(s\bar{q}\bigr)^{0}_{\textbf{8}}\biggr \rangle ^{0}_{\textbf{1}}  \biggr) \\ 
			&-\frac{1}{4} \biggr(\sqrt{\frac{2}{3}}\biggr|\bigr(s\bar{s}\bigr)^{1}_{\textbf{1}}\bigr(q\bar{q}\bigr)^{1}_{\textbf{1}}\biggr \rangle ^{0}_{\textbf{1}}  +\sqrt{\frac{1}{3}}\biggr|\bigr(s\bar{s}\bigr)^{1}_{\textbf{8}}\bigr(q\bar{q}\bigr)^{1}_{\textbf{8}}\biggr \rangle ^{0}_{\textbf{1}}  \biggr)\\
			&+\frac{\sqrt{3}}{4} \biggr(\sqrt{\frac{2}{3}}\biggr|\bigr(s\bar{s}\bigr)^{0}_{\textbf{1}}\bigr(q\bar{q}\bigr)^{0}_{\textbf{1}}\biggr \rangle ^{0}_{\textbf{1}}  +\sqrt{\frac{1}{3}}\biggr|\bigr(s\bar{s}\bigr)^{0}_{\textbf{8}}\bigr(q\bar{q}\bigr)^{0}_{\textbf{8}}\biggr \rangle ^{0}_{\textbf{1}}  \biggr)
		\end{split}
		\label{eq251}
	\end{eqnarray}
	
	where the color representation's dimensions are denoted using subscripts and the total spin is denoted by a superscript. 
	
	\begin{figure*}[t]
		\centering
		\begin{subfigure}{0.32\textwidth}
			\includegraphics[width=0.98\linewidth, height=0.35\textheight]{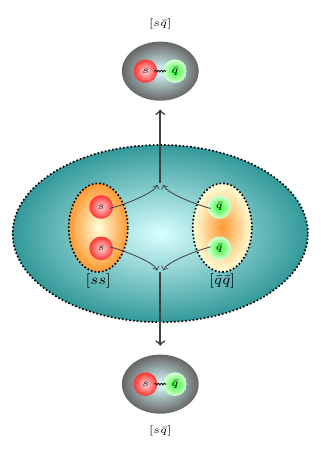}
			\caption{$T_{ss\bar{q}\bar{q}}\rightarrow K + K$}
			\label{fig:tikz---rearrange}
		\end{subfigure}
		\begin{subfigure}{0.32\textwidth}
			\includegraphics[width=0.98\linewidth, height=0.35\textheight]{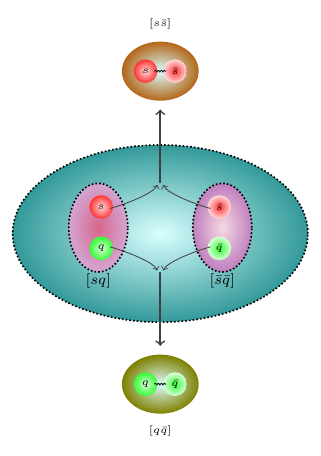}
			\caption{$T_{sq\bar{s}\bar{q}}\rightarrow \phi+\pi$}
			\label{fig:tikz---rearrange1}
		\end{subfigure}
		\begin{subfigure}{0.32\textwidth}
			\includegraphics[width=0.98\linewidth, height=0.35\textheight]{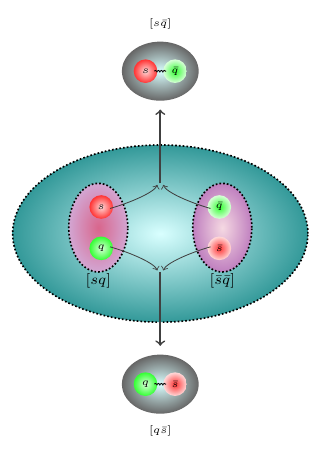}
			\caption{$T_{sq\bar{s}\bar{q}}\rightarrow K + \bar{K}$}
			\label{fig:tikz---rearrange2}
		\end{subfigure}
		\caption{Pictorial representation of tetraquark re-arrangement}
		\label{represent}
	\end{figure*}
	
	The $sq\bar{s}\bar{q}$ tetraquark can be rearranged in two possible manners, namely, $q\bar{s}+s\bar{q}$ and $s\bar{s}+q\bar{q}$ while $ss\bar{q}\bar{q}$ can only rearrange as $q\bar{s}+s\bar{q}$. The decay of various quark-antiquark pairs ($s\bar{s}$,$s\bar{q}$,$q\bar{q}$) into lower mass states leads the description of the rearrangement decay. These decays are seen in various channels, which are as follows:
	
	1. The conversion and confinement of two gluons into light hadrons is obtained by the decay of a color singlet spin-0 $s\bar{s}$ pair with a rate of order of $\alpha_{s}^{2}$. Similarly, the $s\bar{s}$ pair can also convert into 2 photons. With the spectator $q\bar{q}$ pair in account, the decays observed are:
	\begin{center}
		$sq\bar{s}\bar{q}\rightarrow\pi+\text{light hadrons} $ \\
		$sq\bar{s}\bar{q}\rightarrow\pi+2\gamma $
	\end{center}
	The $q\bar{q}$ pair($\pi^{\pm}$) can decay into $e^{\pm}+\nu_{e}$ and $\mu^{\pm}+\nu_{\mu}$ with a rate of order of $\alpha^{2}$.With the spectator $s\bar{s}$ pair in account, the decay observed is:
	\begin{center}
		$sq\bar{s}\bar{q}\rightarrow\eta_{s}+e^{\pm}+\nu_{e}$ \\
		$sq\bar{s}\bar{q}\rightarrow\eta_{s}+\mu^{\pm}+\nu_{\mu}$
	\end{center}
2. The conversion and confinement of three gluons into light hadrons are obtained by the decay of color singlet spin-1 pairs with a rate of order of $\alpha_{s}^{3}$. Successfully, the final states of $\phi+ e^{+}e^{-}$, $\phi+ \mu^{+}\mu^{-}$, $\rho+ e^{+}e^{-}$ and $\rho+ \mu^{+}\mu^{-}$ are obsevered with rates of order $\alpha_{s}^{2}$ by annihilating the light hadron into one photon. With the spectator $\phi$ and $\pi$ mesons in account, the decays observed are:
	\begin{center}
	$sq\bar{s}\bar{q}\rightarrow\phi+\text{light hadrons}$ \\
	$sq\bar{s}\bar{q}\rightarrow\rho+\text{light hadrons}$ \\
	\end{center}
	3. The spin-1 $s\bar{s}$ pair in the color octet representation annihilates into one gluon, which materializes into a pair of up or down quarks, which later recombines with the spectator $q\bar{q}$ pair to produce pions, $\pi^{\pm}$, with a rate of order of $\alpha_{s}^{2}$. The pions further can decay into $\mu^{+}+ \nu_{\mu}$ and $e^{+}+ \nu_{e}$. The decay observed is:
	\begin{center}
	$sq\bar{s}\bar{q}\rightarrow\pi+\pi$
	\end{center}
	4. The spin-0 $s\bar{s}$ pair in color octet representation annihilates into two gluons, which produce a pair of up or down quarks to neutralize the color of the spectator pair with a rate of order of $\alpha_{s}^{2}$ \cite{Luo:2015yio}. The decay observed is:
	\begin{center}
	$sq\bar{s}\bar{q}\rightarrow \rho + (2g)$
	\end{center}
	
	5. The spin-1 $q\bar{s}$ pair in color singlet representation decays into leptonic channels $e^{+}+\nu_{e}$ and $\mu^{+}+\nu_{\mu}$ when q is up quark. Similarly, the spin-1 $s\bar{q}$ pair decays into leptonic channels $e^{-}+\nu_{e}$ and $\mu^{-}+\nu_{\mu}$. This conversion occurs with a rate of order of $\alpha^{2}$. With the spectator $q\bar{s}$ and $s\bar{q}$ pairs in account, the decay observed is:
	\begin{center}
	$sq\bar{s}\bar{q}/ss\bar{q}\bar{q}\rightarrow K^{\pm} + e^{\mp}\nu_{e}$\\
	$sq\bar{s}\bar{q}/ss\bar{q}\bar{q}\rightarrow K^{\pm} + \mu^{\mp}\nu_{\mu}$\\
	\end{center}
	
	6. The spin-1 $q\bar{s}$ pair in color singlet representation can also decay into the rare leptonic channels $\gamma e^{+}+\nu_{e}$ and $\gamma \mu^{+}+\nu_{\mu}$ when q is up quark. Similarly, the spin-1 $s\bar{q}$ pair decays into the rare leptonic channels $\gamma e^{-}+\nu_{e}$ and $\gamma \mu^{-}+\nu_{\mu}$. This conversion occurs with a rate of order of $\alpha^{2}$. With the spectator $q\bar{s}$ and $s\bar{q}$ pairs in account, the decay observed is:
	\begin{center}
	$sq\bar{s}\bar{q}/ss\bar{q}\bar{q}\rightarrow K^{\pm} + \gamma e^{\mp}\nu_{e}$\\
	$sq\bar{s}\bar{q}/ss\bar{q}\bar{q}\rightarrow K^{\pm} + \gamma\mu^{\mp}\nu_{\mu}$\\
	\end{center}

	The ratio of overlap probabilities of the annihilating $s\bar{s}$ and $q\bar{q}$ pair in $sq\bar{s}\bar{q}$ is proportional to the decay rates:
	\begin{equation}
	\varrho_{1} = \frac{|\Psi_{sq\bar{s}\bar{q}}(0)|^{2}}{|\Psi_{\phi}(0)|^{2}}\\
	\varrho_{2} = \frac{|\Psi_{sq\bar{s}\bar{q}}(0)|^{2}}{|\Psi_{\rho}(0)|^{2}}
	\end{equation}
	The ratio of overlap probabilities of the decaying $q\bar{s}$ and $s\bar{q}$ pair in $sq\bar{s}\bar{q}/ss\bar{q}\bar{q}$ is proportional to the decay rates:
	\begin{equation}
	\varrho_{3} = \frac{|\Psi_{sq\bar{s}\bar{q}}(0)|^{2}}{|\Psi_{q\bar{s}/s\bar{q}}(0)|^{2}}\\
	\varrho_{4} = \frac{|\Psi_{{ss\bar{q}\bar{q}}(0)|^{2}}}{|\Psi_{q\bar{s}/s\bar{q}}(0)|^{2}}\\
	\end{equation}
	
	The individual decay rate is obtained using the simple formula \cite{Berestetskii:1982qgu}:
	\begin{equation}
	\Gamma(i)^{spin}_{color}=||\Psi_{sq\bar{s}\bar{q}/ss\bar{q}\bar{q}}(0)|^{2}v\sigma(i)^{spin}_{color}\rightarrow f)
	\end{equation}
	where $|\Psi_{{sq\bar{s}\bar{q}/ss\bar{q}\bar{q}}}(0)|^{2},v,$ and $\sigma$ are the overlap probability of the annihilating or decaying pair, relative velocity, and the spin-averaged annihilation/decay cross section in the final state $f$, respectively. Our previous works \cite{Lodha:2024bwn,Lodha:2024qby} and ref. \cite{Becchi:2020mjz,Becchi:2020uvq} explores the physics of rearrangement decay in further detail. The total of all the individual decay rates yields the total decay rate. The spectator pairs $s\bar{s}$ can appear as $\eta_{s}$ or $\phi$ on the mass shell, or combine with the outgoing $q\bar{q}$ into a pair of open-strange particles for the given tetraquark. Similarly, the spectator pair $q\bar{q}$ can appear as $\eta$ or $\pi$ on the mass shell, or combine with the outgoing $s\bar{s}$ into a pair of open-strange particles. Figure \ref{reaarange} depicts the rearrangement diagram for Tetraquarks $T_{sq\bar{s}\bar{q}}$ and $T_{ss\bar{q}\bar{q}}$. Similarly, figure \ref{represent} shows a pictorial representation of the rearrangement of Tetraquarks $T_{sq\bar{s}\bar{q}}$ and $T_{ss\bar{q}\bar{q}}$.
	
	\subsection*{A. $\bar{\textbf{3}}-\textbf{3}$ Tetraquark}
	
	The singlet spin 0 decay rate is given by:
	
	\begin{equation}
		\begin{split}
			\Gamma_{1}&=\Gamma(sq\bar{s}\bar{q}\rightarrow\pi+\text{light hadrons}	)\\
			&=2\cdot\frac{1}{16}\cdot|\Psi_{sq\bar{s}\bar{q}}(0)|^{2} v \sigma((s\bar{s}^{0}_{1})\rightarrow \text{2 gluons}) \\
			& = \frac{1}{8}\Gamma(\eta_{s})\cdot\varrho_{1} = 0.024 \;\; MeV \cdot\varrho_{1}
		\end{split}
	\end{equation}
	
	\begin{equation}
		\begin{split}
			\Gamma_{2}&=\Gamma(sq\bar{s}\bar{q}\rightarrow\pi+ 2\gamma)\\
			&=2\cdot\frac{1}{16}\cdot|\Psi_{sq\bar{s}\bar{q}}(0)|^{2} v \sigma((s\bar{s}^{0}_{1})\rightarrow 2 \gamma) \\
			& = \frac{1}{8}\Gamma(\eta_{s}\rightarrow2\gamma)\cdot\varrho_{1} = 0.094 \;\; keV \cdot\varrho_{1}
		\end{split}
	\end{equation}

	\begin{equation}
		\begin{split}
			\Gamma_{3}&=\Gamma(sq\bar{s}\bar{q}\rightarrow\eta_{s}+e^{\pm}+\nu_{e})\\
			&=2\cdot\frac{1}{16}\cdot|\Psi_{sq\bar{s}\bar{q}}(0)|^{2} v \sigma((q\bar{q}^{0}_{1})\rightarrow e^{\pm}+\nu_{e}) \\
			& = \frac{1}{8}\Gamma(\pi\rightarrow e^{\pm}+\nu_{e})\cdot\varrho_{2} =  \;\;5.906 \; eV \cdot\varrho_{2}
		\end{split}
	\end{equation}

	\begin{equation}
		\begin{split}
			\Gamma_{4}&=\Gamma(sq\bar{s}\bar{q}\rightarrow\eta_{s}+\mu^{\pm}+\nu_{\mu})\\
			&=2\cdot\frac{1}{16}\cdot|\Psi_{sq\bar{s}\bar{q}}(0)|^{2} v \sigma((q\bar{q}^{0}_{1})\rightarrow \mu^{\pm}+\nu_{\mu}) \\
			& = \frac{1}{8}\Gamma(\pi\rightarrow \mu^{\pm}+\nu_{\mu})\cdot\varrho_{2} = 4.797 \;\; MeV \cdot\varrho_{2}
		\end{split}
	\end{equation}
	
	Similarly, the color singlet spin 1 decay rate is given by
	
	\begin{equation}
		\begin{split}
			\Gamma_{5}&=\Gamma(sq\bar{s}\bar{q} \rightarrow \phi + \text{light hadrons})\\
			&=2\cdot\frac{1}{48}\cdot|\Psi_{sq\bar{s}\bar{q}}(0)|^{2} v \sigma((q\bar{q}^{1}_{1})\rightarrow \text{3 gluons}) \\
			& = \frac{1}{24}\Gamma(\rho)\cdot\varrho_{2} = 6.142 \;\; MeV \cdot\varrho_{2}
		\end{split}
	\end{equation}
	
	\begin{equation}
		\begin{split}
			\Gamma_{6}&=\Gamma(sq\bar{s}\bar{q} \rightarrow \rho + \text{light hadrons})\\
			&=2\cdot\frac{1}{48}\cdot|\Psi_{sq\bar{s}\bar{q}}(0)|^{2} v \sigma((s\bar{s}^{1}_{1})\rightarrow \text{3 gluons}) \\
			& = \frac{1}{24}\Gamma(\phi)\cdot\varrho_{1} = 0.177 \;\; MeV \cdot\varrho_{1}
		\end{split}
	\end{equation}
	
	The annihilation of the spin-1 octet $s\bar{s}$ pair into the light quark pair is given by:
	
	\begin{equation}
		\begin{split}
			\Gamma_{7}=& 2 \cdot\frac{1}{24} \cdot\frac{1}{4}\biggr( \frac{4\pi\alpha_{s}^{2}}{3}\frac{4}{m_{\phi}^{2}} \biggr)|\Psi_{\phi}(0)|^{2} \cdot\varrho_{1}\\
			& = 0.014 \; MeV \cdot\varrho_{1}
		\end{split}
	\end{equation}
	The annihilation of the spin-0 octet $s\bar{s}$ pair into the light quark pair is given by:
	\begin{equation}
		\begin{split}
			\Gamma_{8}=& 2 \cdot\frac{1}{8} \cdot\frac{1}{4}\biggr( \frac{4\pi\alpha_{s}^{2}}{m_{\phi}^{2}}\cdot18 \biggr)|\Psi_{\phi}(0)|^{2} \cdot\varrho_{1}\\
			& = 0.552 \; MeV \cdot\varrho_{1}
		\end{split}
	\end{equation}
	
	The leptonic decay of color singlet spin-1 $u\bar{s}$ pair is given by : 
	
	\begin{equation}
		\begin{split}
			\Gamma_{9}&=\Gamma(sq\bar{s}\bar{q} \rightarrow K^{\pm} + e^{\mp}\nu_{e})\\
			&=2\cdot\frac{1}{48}\cdot|\Psi_{sq\bar{s}\bar{q}}(0)|^{2} v \sigma((K^{\pm})\rightarrow e^{\mp}\nu_{e}) \\
			&=\frac{1}{24} \frac{G_{F}^{2}}{8m_{K^{\pm}}^{3}\pi}f^{2}_{K^{\pm}}[m_{e}(m_{K^{\pm}}^{2}-m_{e}^{2})]^{2} \\
			& = 53.243 \;\; eV \cdot\varrho_{3}
		\end{split}
	\end{equation}

	\begin{equation}
		\begin{split}
			\Gamma_{10}&=\Gamma(sq\bar{s}\bar{q} \rightarrow K^{\pm} + \mu^{\mp}\nu_{\mu})\\
			&=2\cdot\frac{1}{48}\cdot|\Psi_{sq\bar{s}\bar{q}}(0)|^{2} v \sigma(K^{\pm}\rightarrow \mu^{\mp}\nu_{\mu}) \\
			&=\frac{1}{24} \frac{G_{F}^{2}}{8m_{K^{\pm}}^{3}\pi}f^{2}_{K^{\pm}}[m_{\mu}(m_{K^{\pm}}^{2}-m_{\mu}^{2})]^{2} \\
			& =  2.139 \;\; MeV \cdot\varrho_{3}
		\end{split}
	\end{equation}
	
	\begin{equation}
		\begin{split}
			\Gamma_{11}&=\Gamma(ss\bar{q}\bar{q} \rightarrow K^{\pm} + e^{\mp}\nu_{e})\\
			&=2\cdot\frac{1}{12}\cdot|\Psi_{ss\bar{q}\bar{q}}(0)|^{2} v \sigma((K^{\pm})\rightarrow e^{\mp}\nu_{e}) \\
			&=\frac{1}{6} \frac{G_{F}^{2}}{8m_{K^{\pm}}^{3}\pi}f^{2}_{K^{\pm}}[m_{e}(m_{K^{\pm}}^{2}-m_{e}^{2})]^{2} \\
			& = 212.972 \;\; eV \cdot\varrho_{4}
		\end{split}
	\end{equation}

	\begin{equation}
		\begin{split}
			\Gamma_{12}&=\Gamma(ss\bar{q}\bar{q} \rightarrow K^{\pm} + \mu^{\mp}\nu_{\mu})\\
			&=2\cdot\frac{1}{12}\cdot|\Psi_{ss\bar{q}\bar{q}}(0)|^{2} v \sigma(K^{\pm}\rightarrow \mu^{\mp}\nu_{\mu}) \\
			&=\frac{1}{6} \frac{G_{F}^{2}}{8m_{K^{\pm}}^{3}\pi}f^{2}_{K^{\pm}}[m_{\mu}(m_{K^{\pm}}^{2}-m_{\mu}^{2})]^{2} \\
			& = 8.556 \;\; MeV \cdot\varrho_{4}
		\end{split}
	\end{equation}
	
	The rare leptonic decay of color singlet spin-1 $u\bar{s}$ pair is given by : 
	
	\begin{equation}
		\begin{split}
			\Gamma_{13}&=\Gamma(sq\bar{s}\bar{q} \rightarrow K^{\pm} + \gamma e^{\mp}\nu_{e})\\
			&=2\cdot\frac{1}{48}\cdot|\Psi_{sq\bar{s}\bar{q}}(0)|^{2} v \sigma((K^{\pm})\rightarrow \gamma e^{\mp}\nu_{e}) \\
			&=\frac{1}{24}\frac{\alpha G_{F}^{2}}{2592\pi^{2}}|V_{su}|^{2}f^{2}_{K^{\pm}}m_{K^{\pm}}^{3}(x_{u}+x_{s}) \\
			& = 6.395 \;\; eV \cdot\varrho_{3}
		\end{split}
	\end{equation}

	\begin{equation}
		\begin{split}
			\Gamma_{14}&=\Gamma(sq\bar{s}\bar{q} \rightarrow K^{\pm} + \gamma \mu^{\mp}\nu_{\mu})\\
			&=2\cdot\frac{1}{48}\cdot|\Psi_{sq\bar{s}\bar{q}}(0)|^{2} v \sigma(K^{\pm}\rightarrow \gamma \mu^{\mp}\nu_{\mu}) \\
			&=\frac{1}{24}\frac{\alpha G_{F}^{2}}{2592\pi^{2}}|V_{su}|^{2}f^{2}_{K^{\pm}}m_{K^{\pm}}^{3}(x_{u}+x_{s})\\
			& =2.165  \;\; keV \cdot\varrho_{3}
		\end{split}
	\end{equation}
	
	\begin{equation}
		\begin{split}
			\Gamma_{15}&=\Gamma(ss\bar{q}\bar{q} \rightarrow K^{\pm} + \gamma e^{\mp}\nu_{e})\\
			&=2\cdot\frac{1}{12}\cdot|\Psi_{ss\bar{q}\bar{q}}(0)|^{2} v \sigma((K^{\pm})\rightarrow \gamma e^{\mp}\nu_{e}) \\
			&=\frac{1}{6}\frac{\alpha G_{F}^{2}}{2592\pi^{2}}|V_{su}|^{2}f^{2}_{K^{\pm}}m_{K^{\pm}}^{3}(x_{u}+x_{s})\\
			& = 25.578 \;\; eV \cdot\varrho_{4}
		\end{split}
	\end{equation}

	\begin{equation}
		\begin{split}
			\Gamma_{16}&=\Gamma(ss\bar{q}\bar{q} \rightarrow K^{\pm} + \gamma \mu^{\mp}\nu_{\mu})\\
			&=2\cdot\frac{1}{12}\cdot|\Psi_{ss\bar{q}\bar{q}}(0)|^{2} v \sigma(K^{\pm}\rightarrow \gamma \mu^{\mp}\nu_{\mu}) \\
			&=\frac{1}{6}\frac{\alpha G_{F}^{2}}{2592\pi^{2}}|V_{su}|^{2}f^{2}_{K^{\pm}}m_{K^{\pm}}^{3}(x_{u}+x_{s})\\
			& = 8.660 \;\; keV \cdot\varrho_{4}
		\end{split}
	\end{equation}
	where ,
	\begin{equation}
		x_{u} = \biggr(3-\frac{M_{K^{\pm}}}{m_{u}} \biggr)^{2}, \;\;\;\; x_{s} = \biggr(3-2\frac{M_{K^{\pm}}}{m_{s}} \biggr)^{2}.
	\end{equation}
	
	where $G_{F}$ i         s the Fermi constant, $m_{u}$ is the mass of the up quark, $M_{K^{\pm}}$ is the mass of kaon, $f_{K^{\pm}}$ is the weak decay constant, and $|V_{su}|$ is the element of the CKM matrix, respectively. The branching ratio of various decay channels of $T_{sq\bar{s}\bar{q}}$ and $T_{ss\bar{q}\bar{q}}$ tetraquark in antitriplet-triplet configuration have been tabulated in table \ref{spectatordecay}.

	\subsection*{B. $\textbf{6}-\bar{\textbf{6}}$ Tetraquark}
	
	The singlet spin 0 decay rate is given by:
	
	\begin{equation}
		\begin{split}
			\Gamma_{17}&=\Gamma(sq\bar{s}\bar{q}\rightarrow\pi+\text{light hadrons}	)\\
			&=2\cdot\frac{1}{8}\cdot|\Psi_{sq\bar{s}\bar{q}}(0)|^{2} v \sigma((s\bar{s}^{0}_{1})\rightarrow \text{2 gluons}) \\
			& = \frac{1}{4}\Gamma(\eta_{s})\cdot\varrho_{1} = 0.047 \;\; MeV \cdot\varrho_{1}
		\end{split}
	\end{equation}
	
	\begin{equation}
		\begin{split}
			\Gamma_{18}&=\Gamma(sq\bar{s}\bar{q}\rightarrow\pi+ 2\gamma)\\
			&=2\cdot\frac{1}{8}\cdot|\Psi_{sq\bar{s}\bar{q}}(0)|^{2} v \sigma((s\bar{s}^{0}_{1})\rightarrow 2 \gamma) \\
			& = \frac{1}{4}\Gamma(\eta_{s}\rightarrow2\gamma)\cdot\varrho_{1} = 0.188 \;\; keV \cdot\varrho_{1}
		\end{split}
	\end{equation}

	\begin{equation}
		\begin{split}
			\Gamma_{19}&=\Gamma(sq\bar{s}\bar{q}\rightarrow\eta_{s}+e^{\pm}+\nu_{e})\\
			&=2\cdot\frac{1}{8}\cdot|\Psi_{sq\bar{s}\bar{q}}(0)|^{2} v \sigma((q\bar{q}^{0}_{1})\rightarrow e^{\pm}+\nu_{e}) \\
			& = \frac{1}{4}\Gamma(\pi\rightarrow e^{\pm}+\nu_{e})\cdot\varrho_{2} =  \;\;11.812 \; eV \cdot\varrho_{2}
		\end{split}
	\end{equation}

	\begin{equation}
		\begin{split}
			\Gamma_{20}&=\Gamma(sq\bar{s}\bar{q}\rightarrow\eta_{s}+\mu^{\pm}+\nu_{\mu})\\
			&=2\cdot\frac{1}{8}\cdot|\Psi_{sq\bar{s}\bar{q}}(0)|^{2} v \sigma((q\bar{q}^{0}_{1})\rightarrow \mu^{\pm}+\nu_{\mu}) \\
			& = \frac{1}{4}\Gamma(\pi\rightarrow \mu^{\pm}+\nu_{\mu})\cdot\varrho_{2} = 9.594 \;\; MeV \cdot\varrho_{2}
		\end{split}
	\end{equation}
	
	Similarly, the color singlet spin 1 decay rate is given by
	
	\begin{equation}
		\begin{split}
			\Gamma_{21}&=\Gamma(sq\bar{s}\bar{q} \rightarrow \phi + \text{light hadrons})\\
			&=2\cdot\frac{1}{24}\cdot|\Psi_{sq\bar{s}\bar{q}}(0)|^{2} v \sigma((q\bar{q}^{1}_{1})\rightarrow \text{3 gluons}) \\
			& = \frac{1}{12}\Gamma(\rho)\cdot\varrho_{2} = 6.142 \;\; MeV \cdot\varrho_{2}
		\end{split}
	\end{equation}
	
	\begin{equation}
		\begin{split}
			\Gamma_{22}&=\Gamma(sq\bar{s}\bar{q} \rightarrow \rho + \text{light hadrons})\\
			&=2\cdot\frac{1}{24}\cdot|\Psi_{sq\bar{s}\bar{q}}(0)|^{2} v \sigma((s\bar{s}^{1}_{1})\rightarrow \text{3 gluons}) \\
			& = \frac{1}{12}\Gamma(\phi)\cdot\varrho_{1} = 0.355 \;\; MeV \cdot\varrho_{1}
		\end{split}
	\end{equation}
	
	The annihilation of the spin-1 octet $s\bar{s}$ pair into the light quark pair is given by:
	
	\begin{equation}
		\begin{split}
			\Gamma_{23}=& 2 \cdot\frac{1}{48} \cdot\frac{1}{4}\biggr( \frac{4\pi\alpha_{s}^{2}}{3}\frac{4}{m_{\phi}^{2}} \biggr)|\Psi_{\phi}(0)|^{2} \cdot\varrho_{1}\\
			& = 0.007 \; MeV \cdot\varrho_{1}
		\end{split}
	\end{equation}
	The annihilation of the spin-0 octet $s\bar{s}$ pair into the light quark pair is given by:
	\begin{equation}
		\begin{split}
			\Gamma_{24}=& 2 \cdot\frac{1}{16} \cdot\frac{1}{4}\biggr( \frac{4\pi\alpha_{s}^{2}}{m_{\phi}^{2}}\cdot18 \biggr)|\Psi_{\phi}(0)|^{2} \cdot\varrho_{1}\\
			& = 0.276 \; MeV \cdot\varrho_{1}
		\end{split}
	\end{equation}
	
	The leptonic decay of color singlet spin-1 $u\bar{s}$ pair is given by : 
	
	\begin{equation}
		\begin{split}
			\Gamma_{25}&=\Gamma(sq\bar{s}\bar{q} \rightarrow K^{\pm} + e^{\mp}\nu_{e})\\
			&=2\cdot\frac{1}{24}\cdot|\Psi_{sq\bar{s}\bar{q}}(0)|^{2} v \sigma((K^{\pm})\rightarrow e^{\mp}\nu_{e}) \\
			&=\frac{1}{12} \frac{G_{F}^{2}}{8m_{K^{\pm}}^{3}\pi}f^{2}_{K^{\pm}}[m_{e}(m_{K^{\pm}}^{2}-m_{e}^{2})]^{2} \\
			& = 106.486 \;\; eV \cdot\varrho_{3}
		\end{split}
	\end{equation}

	\begin{equation}
		\begin{split}
			\Gamma_{26}&=\Gamma(sq\bar{s}\bar{q} \rightarrow K^{\pm} + \mu^{\mp}\nu_{\mu})\\
			&=2\cdot\frac{1}{24}\cdot|\Psi_{sq\bar{s}\bar{q}}(0)|^{2} v \sigma(K^{\pm}\rightarrow \mu^{\mp}\nu_{\mu}) \\
			&=\frac{1}{12} \frac{G_{F}^{2}}{8m_{K^{\pm}}^{3}\pi}f^{2}_{K^{\pm}}[m_{\mu}(m_{K^{\pm}}^{2}-m_{\mu}^{2})]^{2} \\
			& =  4.276 \;\; MeV \cdot\varrho_{3}
		\end{split}
	\end{equation}
	
	\begin{equation}
		\begin{split}
			\Gamma_{27}&=\Gamma(ss\bar{q}\bar{q} \rightarrow K^{\pm} + e^{\mp}\nu_{e})\\
			&=2\cdot\frac{1}{4}\cdot|\Psi_{ss\bar{q}\bar{q}}(0)|^{2} v \sigma((K^{\pm})\rightarrow e^{\mp}\nu_{e}) \\
			&=\frac{1}{2} \frac{G_{F}^{2}}{8m_{K^{\pm}}^{3}\pi}f^{2}_{K^{\pm}}[m_{e}(m_{K^{\pm}}^{2}-m_{e}^{2})]^{2} \\
			& = 638.916 \;\; eV \cdot\varrho_{4}
		\end{split}
	\end{equation}

	\begin{equation}
		\begin{split}
			\Gamma_{28}&=\Gamma(ss\bar{q}\bar{q} \rightarrow K^{\pm} + \mu^{\mp}\nu_{\mu})\\
			&=2\cdot\frac{1}{4}\cdot|\Psi_{ss\bar{q}\bar{q}}(0)|^{2} v \sigma(K^{\pm}\rightarrow \mu^{\mp}\nu_{\mu}) \\
			&=\frac{1}{2} \frac{G_{F}^{2}}{8m_{K^{\pm}}^{3}\pi}f^{2}_{K^{\pm}}[m_{\mu}(m_{K^{\pm}}^{2}-m_{\mu}^{2})]^{2} \\
			& = 25.668 \;\; MeV \cdot\varrho_{4}
		\end{split}
	\end{equation}
	
	The rare leptonic decay of color singlet spin-1 $u\bar{s}$ pair is given by : 
	
	\begin{equation}
		\begin{split}
			\Gamma_{29}&=\Gamma(sq\bar{s}\bar{q} \rightarrow K^{\pm} + \gamma e^{\mp}\nu_{e})\\
			&=2\cdot\frac{1}{24}\cdot|\Psi_{sq\bar{s}\bar{q}}(0)|^{2} v \sigma((K^{\pm})\rightarrow \gamma e^{\mp}\nu_{e}) \\
			&=\frac{1}{12}\frac{\alpha G_{F}^{2}}{2592\pi^{2}}|V_{su}|^{2}f^{2}_{K^{\pm}}m_{K^{\pm}}^{3}(x_{u}+x_{s}) \\
			& = 12.791 \;\; eV \cdot\varrho_{3}
		\end{split}
	\end{equation}

	\begin{equation}
		\begin{split}
			\Gamma_{30}&=\Gamma(sq\bar{s}\bar{q} \rightarrow K^{\pm} + \gamma \mu^{\mp}\nu_{\mu})\\
			&=2\cdot\frac{1}{24}\cdot|\Psi_{sq\bar{s}\bar{q}}(0)|^{2} v \sigma(K^{\pm}\rightarrow \gamma \mu^{\mp}\nu_{\mu}) \\
			&=\frac{1}{12}\frac{\alpha G_{F}^{2}}{2592\pi^{2}}|V_{su}|^{2}f^{2}_{K^{\pm}}m_{K^{\pm}}^{3}(x_{u}+x_{s})\\
			& =4.331  \;\; keV \cdot\varrho_{3}
		\end{split}
	\end{equation}
	
	\begin{equation}
		\begin{split}
			\Gamma_{31}&=\Gamma(ss\bar{q}\bar{q} \rightarrow K^{\pm} + \gamma e^{\mp}\nu_{e})\\
			&=2\cdot\frac{1}{4}\cdot|\Psi_{ss\bar{q}\bar{q}}(0)|^{2} v \sigma((K^{\pm})\rightarrow \gamma e^{\mp}\nu_{e}) \\
			&=\frac{1}{2}\frac{\alpha G_{F}^{2}}{2592\pi^{2}}|V_{su}|^{2}f^{2}_{K^{\pm}}m_{K^{\pm}}^{3}(x_{u}+x_{s})\\
			& = 76.734 \;\; eV \cdot\varrho_{4}
		\end{split}
	\end{equation}

	\begin{equation}
		\begin{split}
			\Gamma_{32}&=\Gamma(ss\bar{q}\bar{q} \rightarrow K^{\pm} + \gamma \mu^{\mp}\nu_{\mu})\\
			&=2\cdot\frac{1}{12}\cdot|\Psi_{ss\bar{q}\bar{q}}(0)|^{2} v \sigma(K^{\pm}\rightarrow \gamma \mu^{\mp}\nu_{\mu}) \\
			&=\frac{1}{6}\frac{\alpha G_{F}^{2}}{2592\pi^{2}}|V_{su}|^{2}f^{2}_{K^{\pm}}m_{K^{\pm}}^{3}(x_{u}+x_{s})\\
			& = 25.98 \;\; keV \cdot\varrho_{4}
		\end{split}
	\end{equation}
	where ,
	\begin{equation}
		x_{u} = \biggr(3-\frac{M_{K^{\pm}}}{m_{u}} \biggr)^{2}, \;\;\;\; x_{s} = \biggr(3-2\frac{M_{K^{\pm}}}{m_{s}} \biggr)^{2}.
	\end{equation}
	
	The branching ratio of various decay channels of $T_{sq\bar{s}\bar{q}}$ and $T_{ss\bar{q}\bar{q}}$ tetraquark in sextet-antisextet configuration have been tabulated in table \ref{spectatordecay}.

	\begin{table*}
		\centering
		\caption{Branching ratio for various decay channels in spectator model for $T_{^{1}S_{0}}$}
		\label{spectatordecay}
		\begin{tabular}{ccccc}
			\hline
			\multirow{2}{*}{State}  &\multicolumn{2}{c}{$\bar{\textbf{3}}-\textbf{3}$} &\multicolumn{2}{c}{$\textbf{6}-\bar{\textbf{6}}$} \\
			& Semi-Relativistic & Non-Relativistic & Semi-Relativistic & Non-Relativistic \\
			\hline
			$sq\bar{s}\bar{q}\rightarrow\pi+\text{ light hadrons}$ &$3.569\times10^{-4}$  &2.38258$\times10^{-4}$ &$8.98094\times10^{-4}$  &$8.84389\times10^{-4}$  \\
			$sq\bar{s}\bar{q}\rightarrow\pi+ 2\gamma$ & $1.398\times10^{-6}$ &$0.932699\times10^{-6}$   &$3.59237\times10^{-6}$  &$3.5375\times10^{-6}$ \\
			$sq\bar{s}\bar{q}\rightarrow\eta_{s}+e^{\pm}+\nu_{e}$ &$8.784\times10^{-8}$  &$5.86186\times10^{-8}$ &$2.933\times10^{-7}$  &$4.624\times10^{-7}$  \\
			$sq\bar{s}\bar{q}\rightarrow\eta_{s}+\mu^{\pm}+\nu_{\mu}$  &0.07134  &0.04772&0.238232  &0.37709  \\
			$sq\bar{s}\bar{q}\rightarrow\phi+\text{ light hadrons}$ &0.09270  &0.080579   &0.1525  &0.2414\\
			$sq\bar{s}\bar{q}\rightarrow\rho+\text{ light hadrons}$ &$0.1187$  &0.103181   &$6.783\times10^{-3}$  &$6.6799\times10^{-3}$\\
			$sq\bar{s}\bar{q}\rightarrow\pi+\pi$ &$2.0823\times10^{-4}$  &$1.38918\times10^{-4}$   &$1.3375\times10^{-4}$  &$1.3178\times10^{-4}$\\
			$sq\bar{s}\bar{q}\rightarrow\rho+2g$ &$8.210\times10^{-3}$  &$5.47858\times10^{-3}$   &$5.2739\times10^{-4}$  &$5.1934\times10^{-4}$\\
			$sq\bar{s}\bar{q}\rightarrow K^{\pm}+e^{\mp}\nu_{e}$ &$7.9073\times10^{-7}$  &$7.90259\times10^{-7}$ &$2.0318\times10^{-6}$  &$2.9966\times10^{-6} $ \\
			$sq\bar{s}\bar{q}\rightarrow K^{\pm}+\mu^{\mp}\nu_{\mu}$  &0.031767  &0.031748&0.081589  &0.12033  \\
			$ss\bar{q}\bar{q}\rightarrow K^{\pm}+e^{\mp}\nu_{e}$ &$2.996\times10^{-6}$  &$3.34439\times10^{-6}$   &$1.142\times10^{-5}$  &$1.854\times10^{-5}$\\
			$ss\bar{q}\bar{q}\rightarrow K^{\pm}+\mu^{\mp}\nu_{\mu}$ &0.12037  &0.13436   &0.4509  &0.7449\\
			$sq\bar{s}\bar{q}\rightarrow K^{\pm}+\gamma e^{\mp}\nu_{e}$ &$9.49745\times10^{-8}$  &$9.49178\times10^{-8}$   &$2.44062\times10^{-7}$  &$3.5995\times10^{-7}$\\
			$sq\bar{s}\bar{q}\rightarrow K^{\pm}+\gamma\mu^{\mp}\nu_{\mu}$ &$3.2153\times10^{-5}$  &$3.2134\times10^{-5}$   &$8.2638\times10^{-5}$  &$1.21881\times10^{-4}$\\
			$ss\bar{q}\bar{q}\rightarrow K^{\pm}+\gamma e^{\mp}\nu_{e}$ &$3.5985\times10^{-7}$  &$4.01661\times10^{-7}$   &$1.3723\times10^{-6} $ &$2.22716\times10^{-6}$\\
			$ss\bar{q}\bar{q}\rightarrow K^{\pm}+\gamma\mu^{\mp}\nu_{\mu}$ &$1.2183\times10^{-4}$  &$1.35381\times10^{-4}$   &$4.64623\times10^{-4}$  &$7.5405\times10^{-4}$\\

			\hline
		\end{tabular}
	\end{table*}

\section{Regge trajectories}
This section discusses regge trajectories of the calculated mass spectrum of Kaon mesons and all strange tetraquarks. Regge trajectories describe the relationship between a particle's spin (or angular momentum, J) and its mass squared $(M^{2})$. These trajectories indicate that particles can be grouped into families, each following a nearly linear path in a J versus $M^{2}$ plot. For light mesons, composed of quarks such as up, down, and strange quarks, the trajectories are almost linear with a slight convexity. This convexity arises from the dynamics of the light quarks and the strong force binding them, where the gluonic field predominantly influences their behavior. The small deviation from linearity is due to relativistic effects and the non-linear nature of the strong interaction at low energies. In contrast, heavy mesons, containing at least one heavy quark, exhibit concave Regge trajectories. The concavity results from the heavier quark masses dominating the system's dynamics over the gluonic field. In these mesons, the energy needed to increase spin does not correspond directly to a proportional increase in mass squared, leading to a downward-curving trajectory. Heavy quarks, moving more slowly and behaving non-relativistically, cause the meson to act more like a rigid rotor than a flexible string, further contributing to this deviation from linearity seen in light mesons \cite{Godfrey:1985xj,Collins:1977jy,Anisovich:2002us}. In the present work, the tetraquark is modeled as a heavy-heavy particle-antiparticle system, following a similar Regge behavior.

\begin{figure*}[t]
	\centering
\begin{subfigure}{0.475\textwidth}
		\includegraphics[width=1.05\linewidth, height=0.3\textheight]{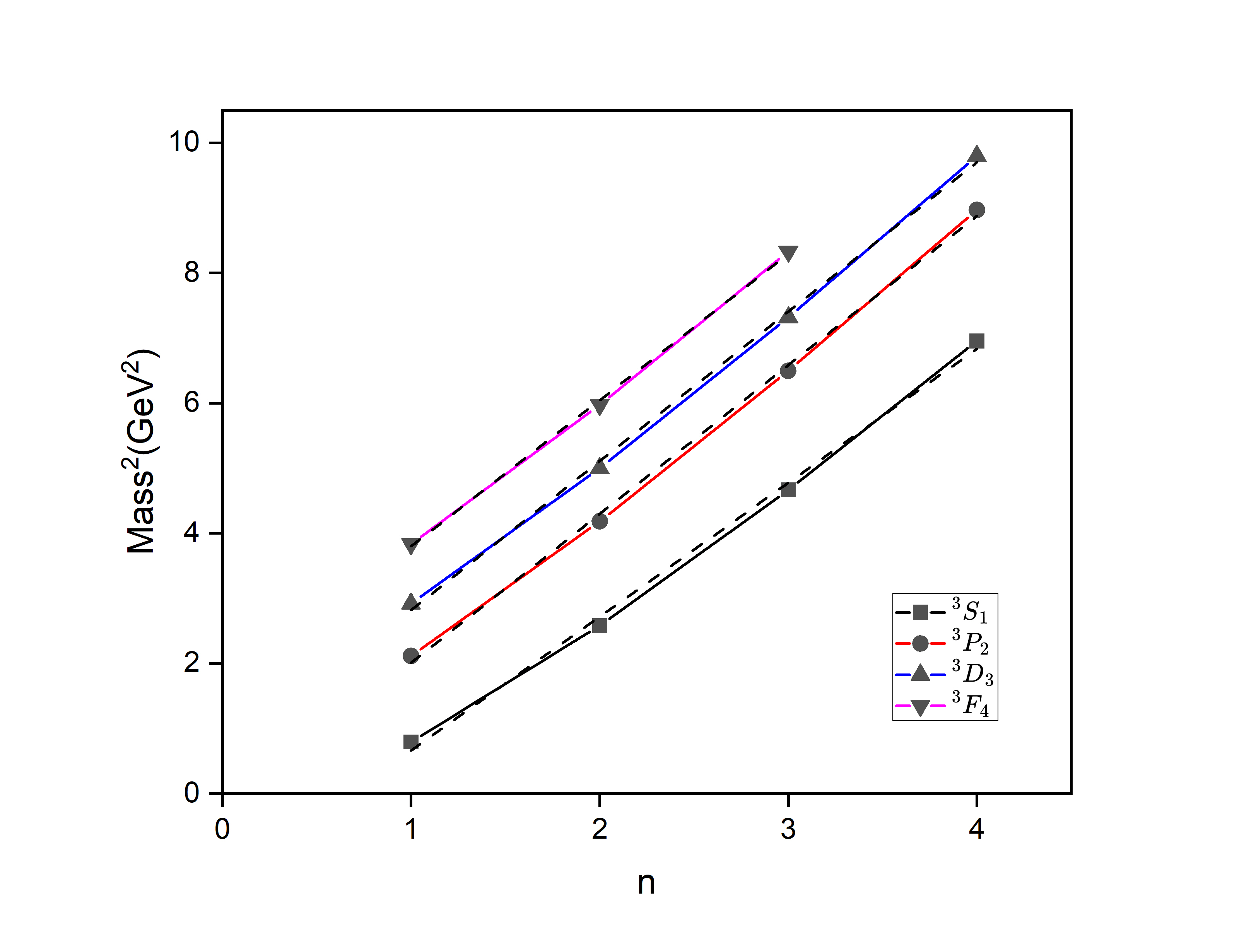}
		\caption{Non-Relativistic formalism}
		\label{fig:mesonnrgraph1nat}
\end{subfigure}
\begin{subfigure}{0.475\textwidth}
		\includegraphics[width=1.05\linewidth, height=0.3\textheight]{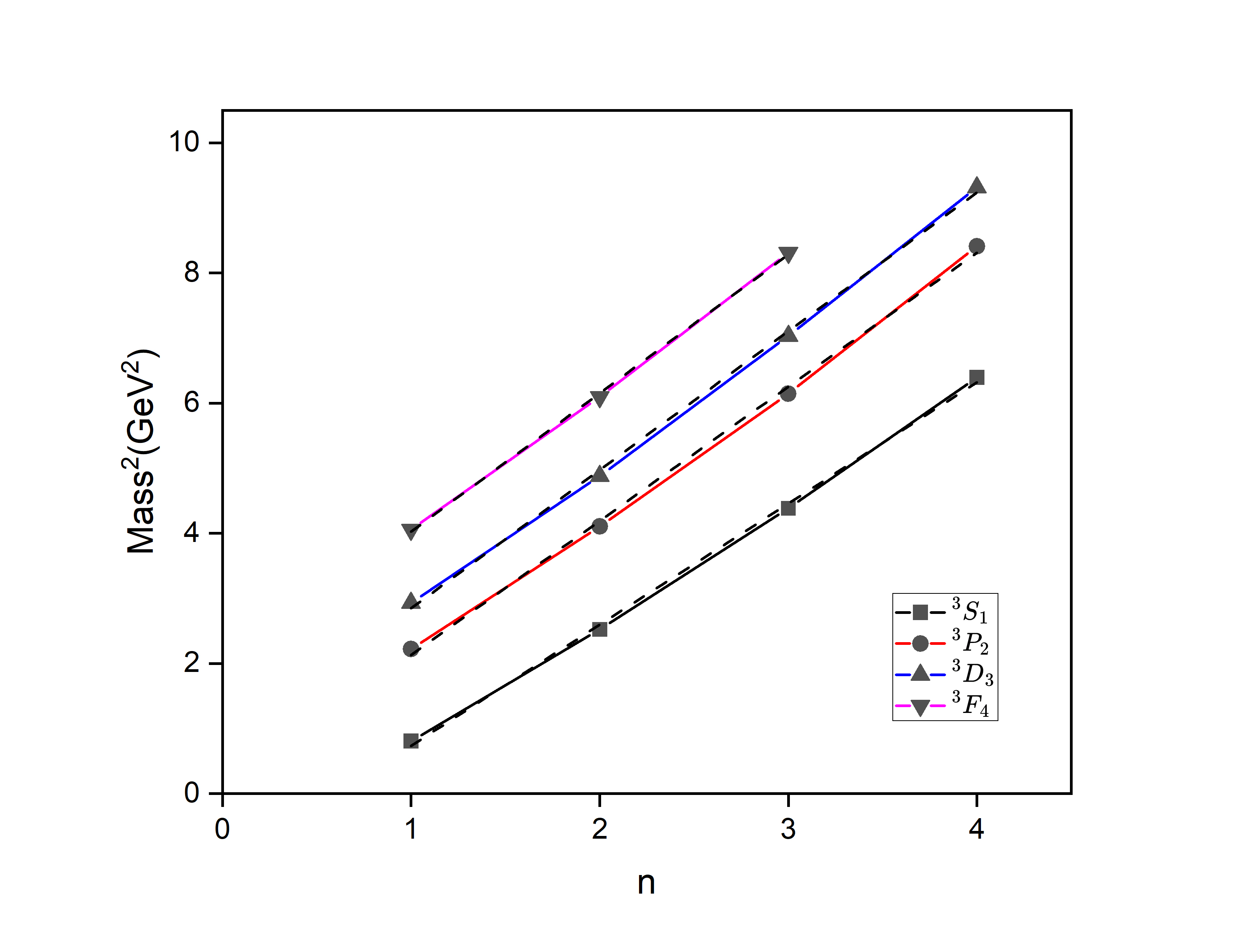}
		\caption{Non-Relativistic formalism}
		\label{fig:mesonsrgraph1nat}
\end{subfigure}

	\caption[]{Regge trajectory in the $(n, M^{2})$ plane for Kaon meson with unnatural parity, (Spin S = 0)}
\end{figure*}

\begin{figure*}[t]
	\centering
	\begin{subfigure}{0.475\textwidth}
		\includegraphics[width=1\linewidth, height=0.3\textheight]{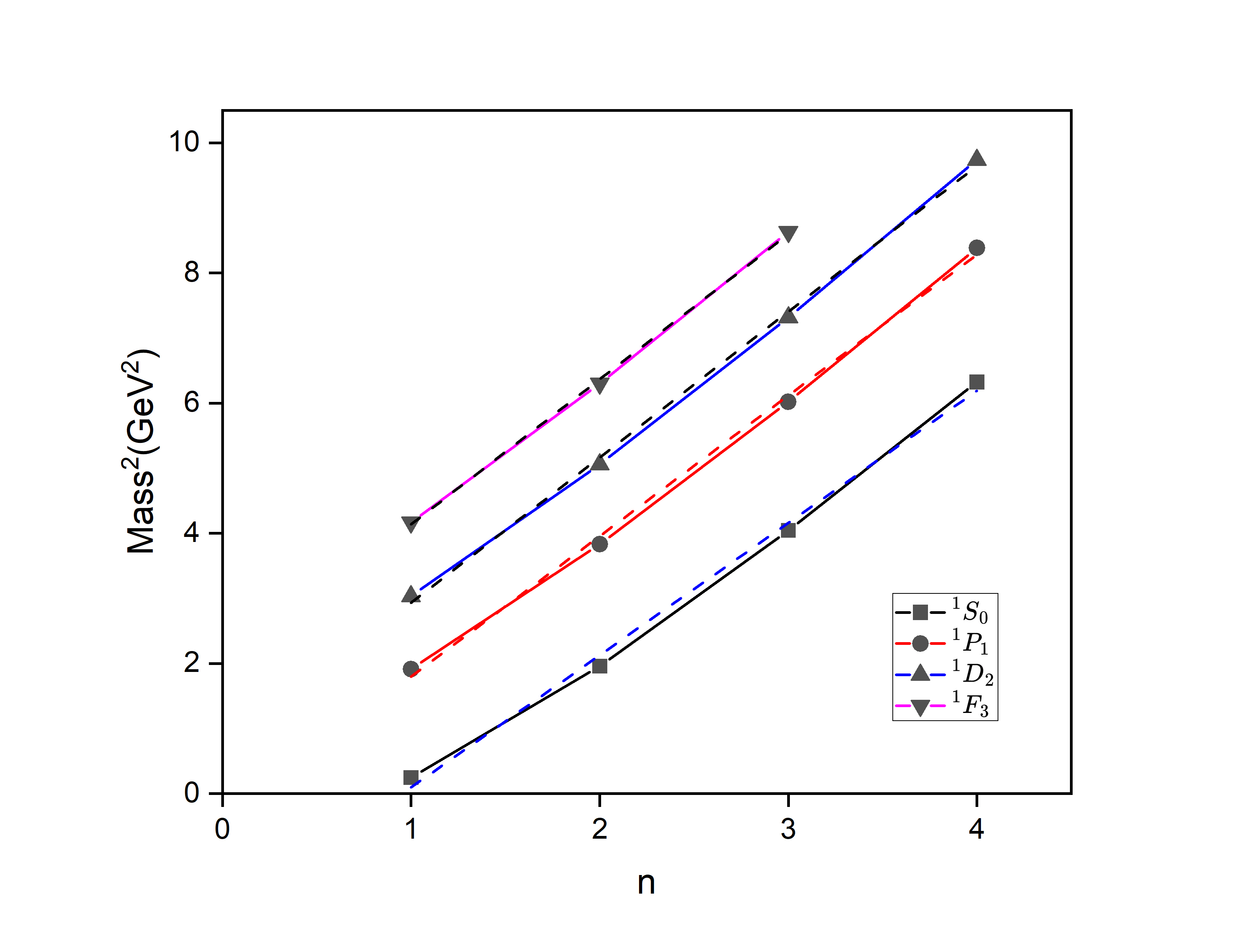}
		\caption{Non-Relativistic formalism}
		\label{fig:mesonnrgraph2nat}
	\end{subfigure}
	\begin{subfigure}{0.475\textwidth}
		\includegraphics[width=1\linewidth, height=0.3\textheight]{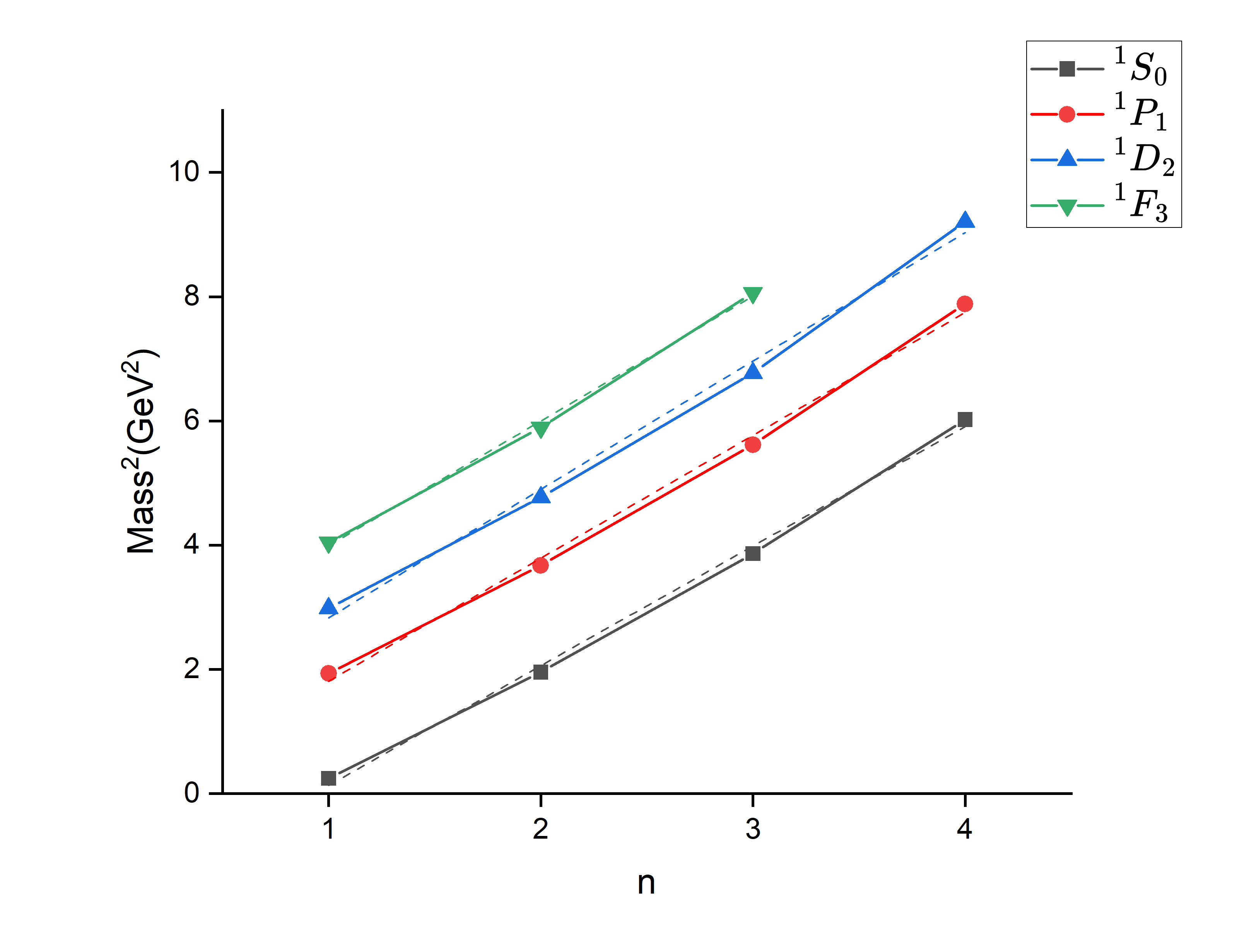}
		\caption{Non-Relativistic formalism}
		\label{fig:mesonsrgraph2nat}
	\end{subfigure}
	\caption[]{Regge trajectory in the $(n, M^{2})$ plane for Kaon meson with natural parity, (Spin S = 1)}
\end{figure*}

\begin{figure*}[t]
	\centering
	\begin{subfigure}{0.475\textwidth}
		\includegraphics[width=1\linewidth, height=0.3\textheight]{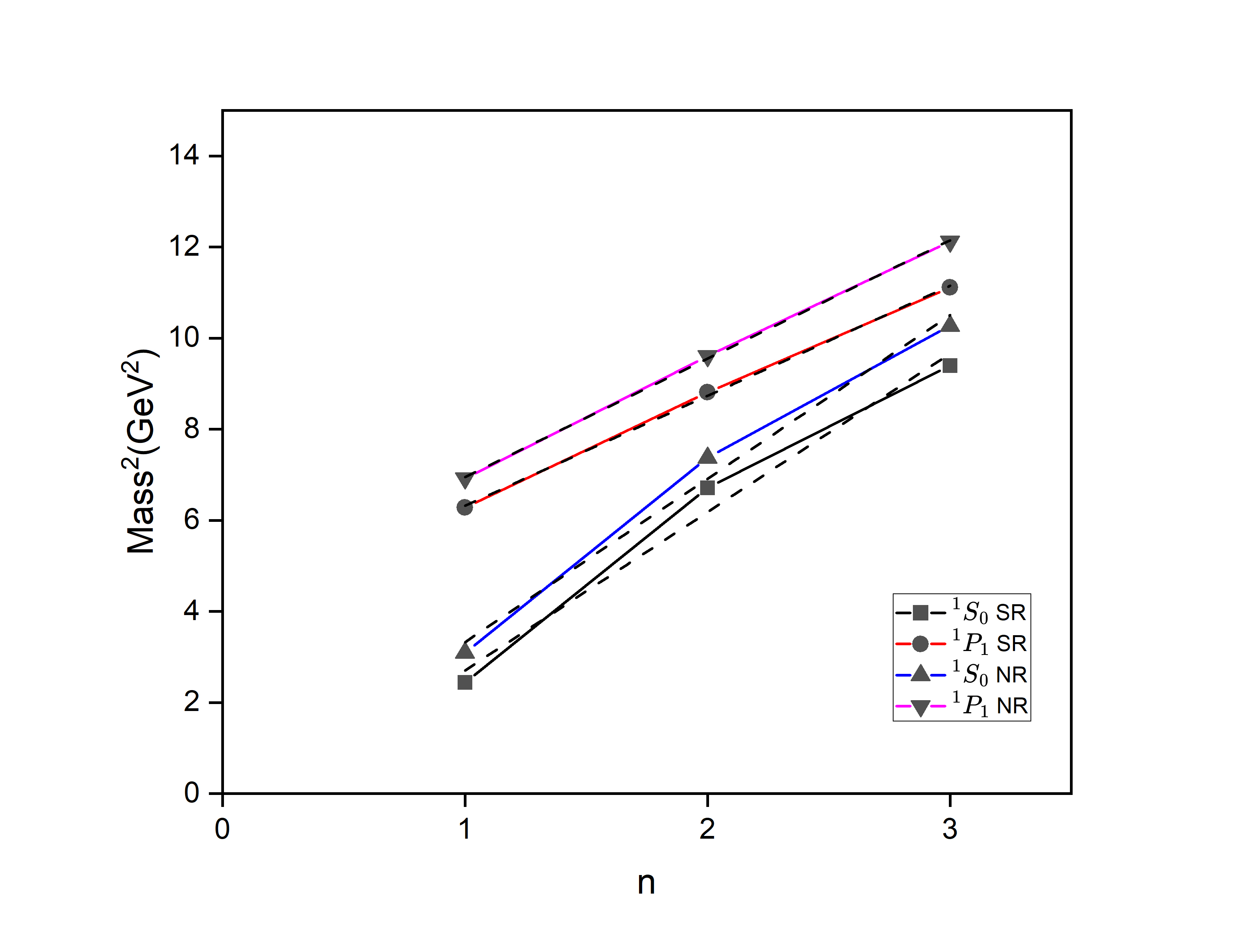}
		\caption{$sq\bar{s}\bar{q}$}
		\label{fig:tetraSQSQS0}
	\end{subfigure}
	\begin{subfigure}{0.475\textwidth}
		\includegraphics[width=1\linewidth, height=0.3\textheight]{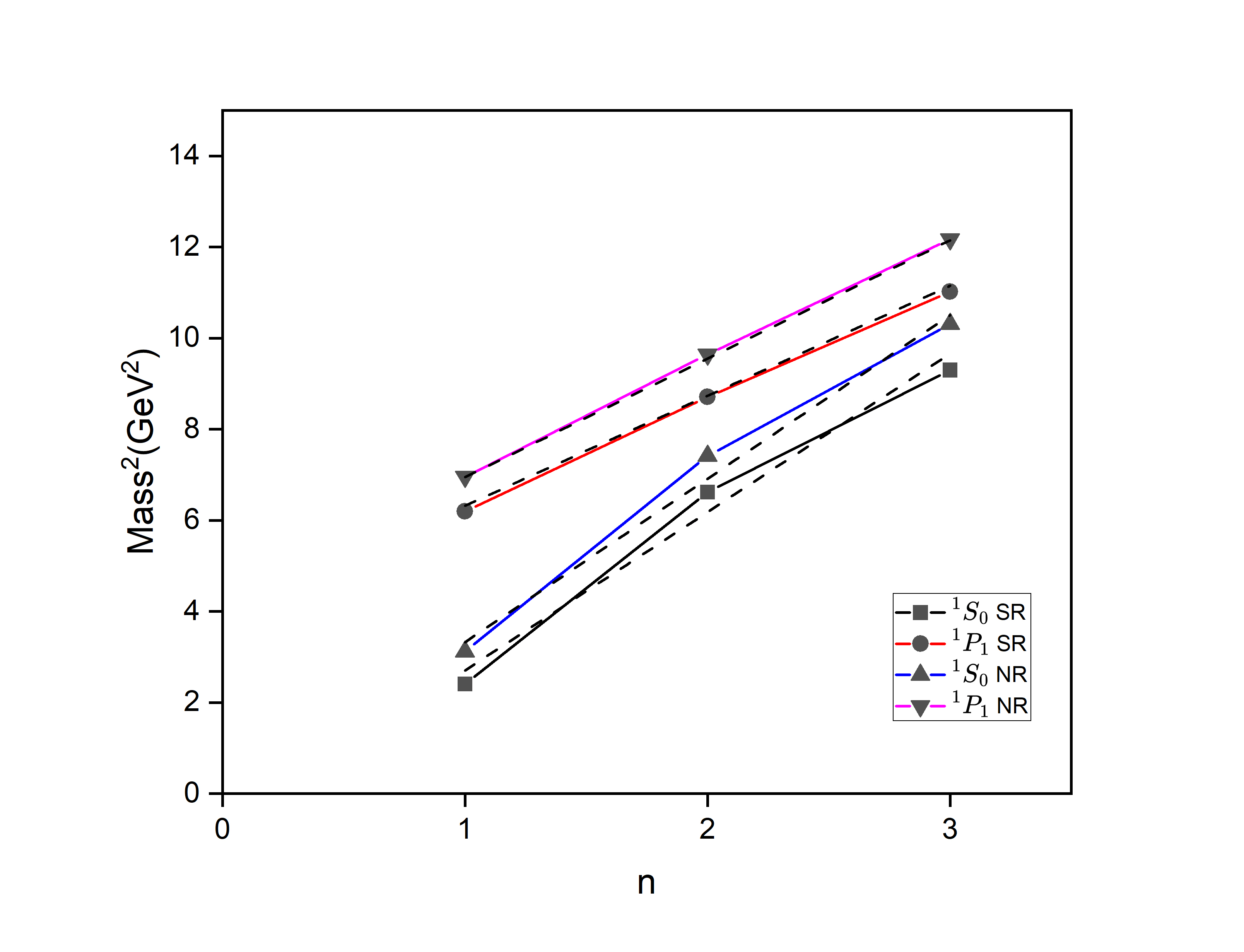}
		\caption{$ss\bar{q}\bar{q}$}
		\label{fig:tetraSSQQ0}
	\end{subfigure}
	\caption[]{Regge trajectory in the $(n, M^{2})$ plane for tetraquarks with Spin S = 0}
\end{figure*}

\begin{figure*}[t]
	\centering
	\begin{subfigure}{0.475\textwidth}
		\includegraphics[width=1\linewidth, height=0.3\textheight]{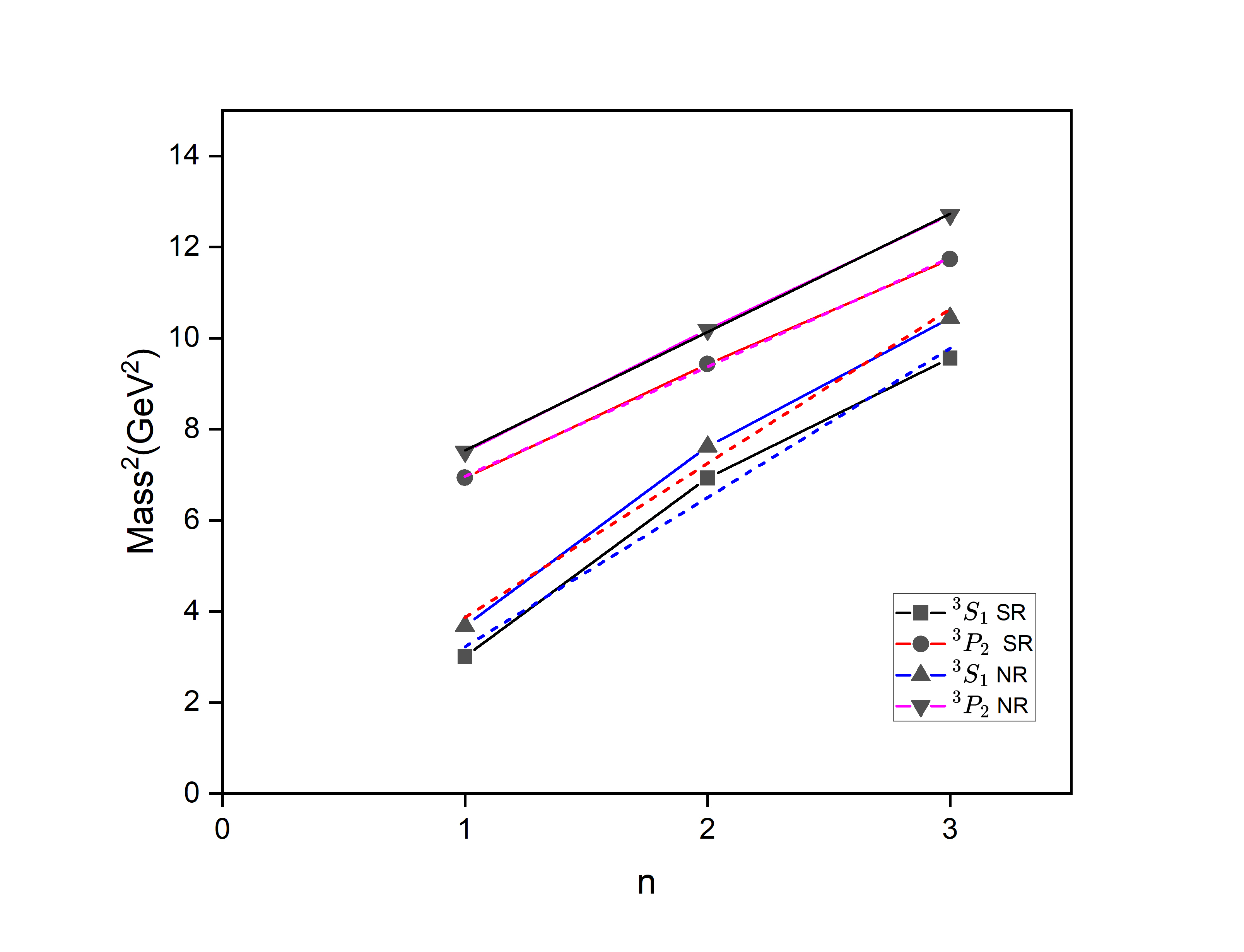}
		\caption{$sq\bar{s}\bar{q}$}
		\label{fig:tetraSQSQS1}
	\end{subfigure}
	\begin{subfigure}{0.475\textwidth}
		\includegraphics[width=1\linewidth, height=0.3\textheight]{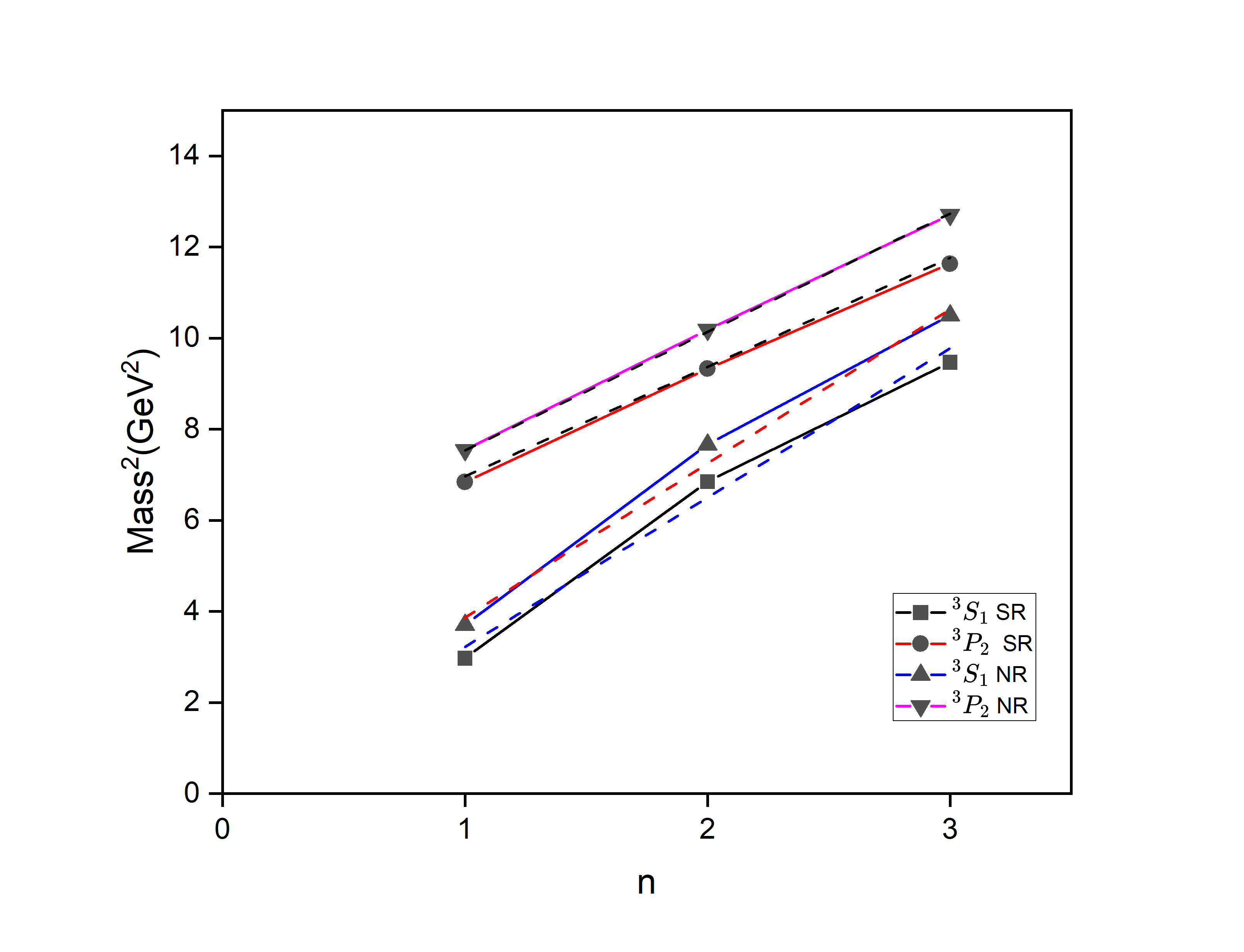}
		\caption{$ss\bar{q}\bar{q}$}
		\label{fig:tetraSSQQ1}
	\end{subfigure}
	\caption[]{Regge trajectory in the $(n, M^{2})$ plane for tetraquarks with Spin S = 1}
\end{figure*}

\begin{figure*}[t]
	\centering
	\begin{subfigure}{0.475\textwidth}
		\includegraphics[width=1\linewidth, height=0.3\textheight]{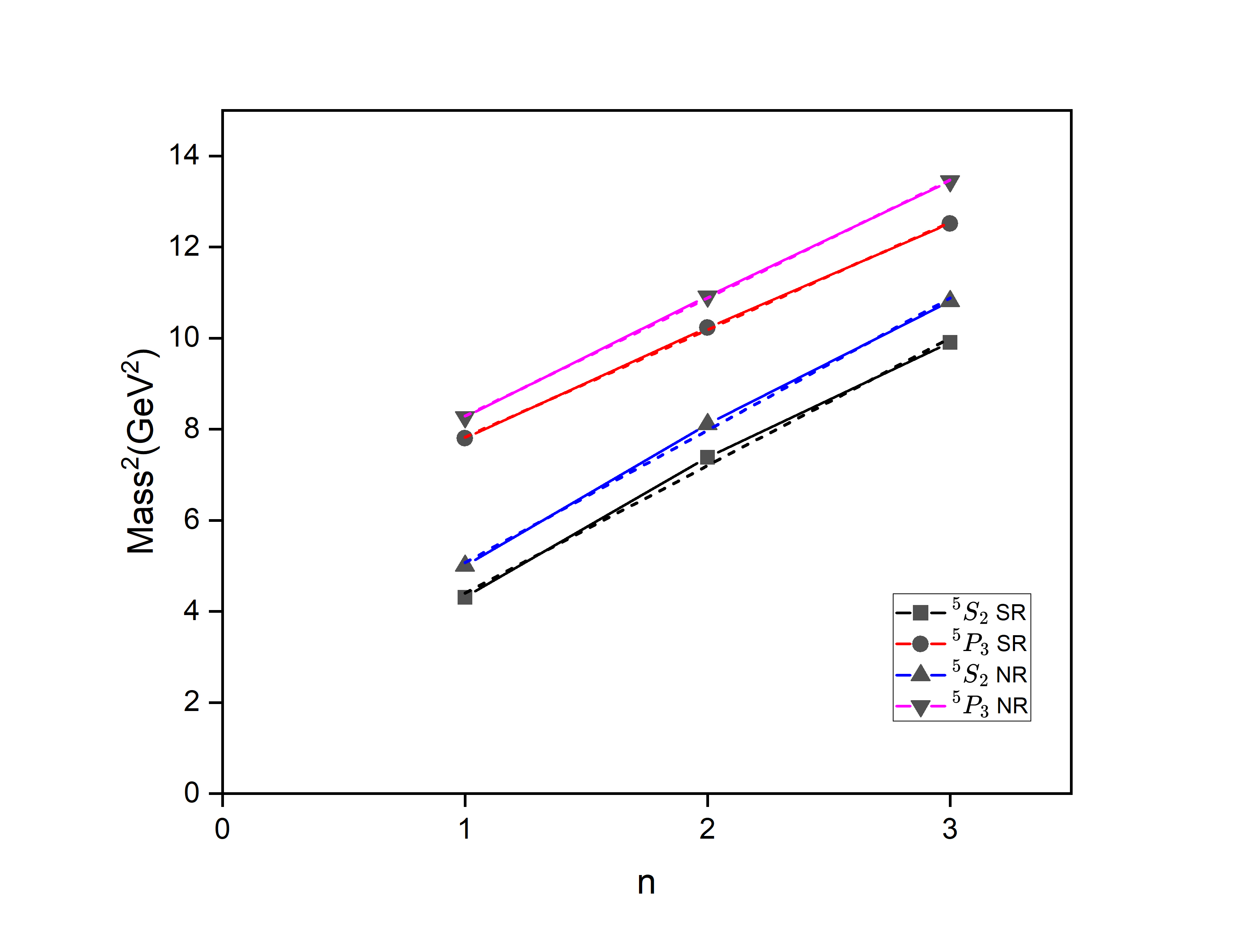}
		\caption{$sq\bar{s}\bar{q}$}
		\label{fig:tetraSQSQS2}
	\end{subfigure}
	\begin{subfigure}{0.475\textwidth}
		\includegraphics[width=1\linewidth, height=0.3\textheight]{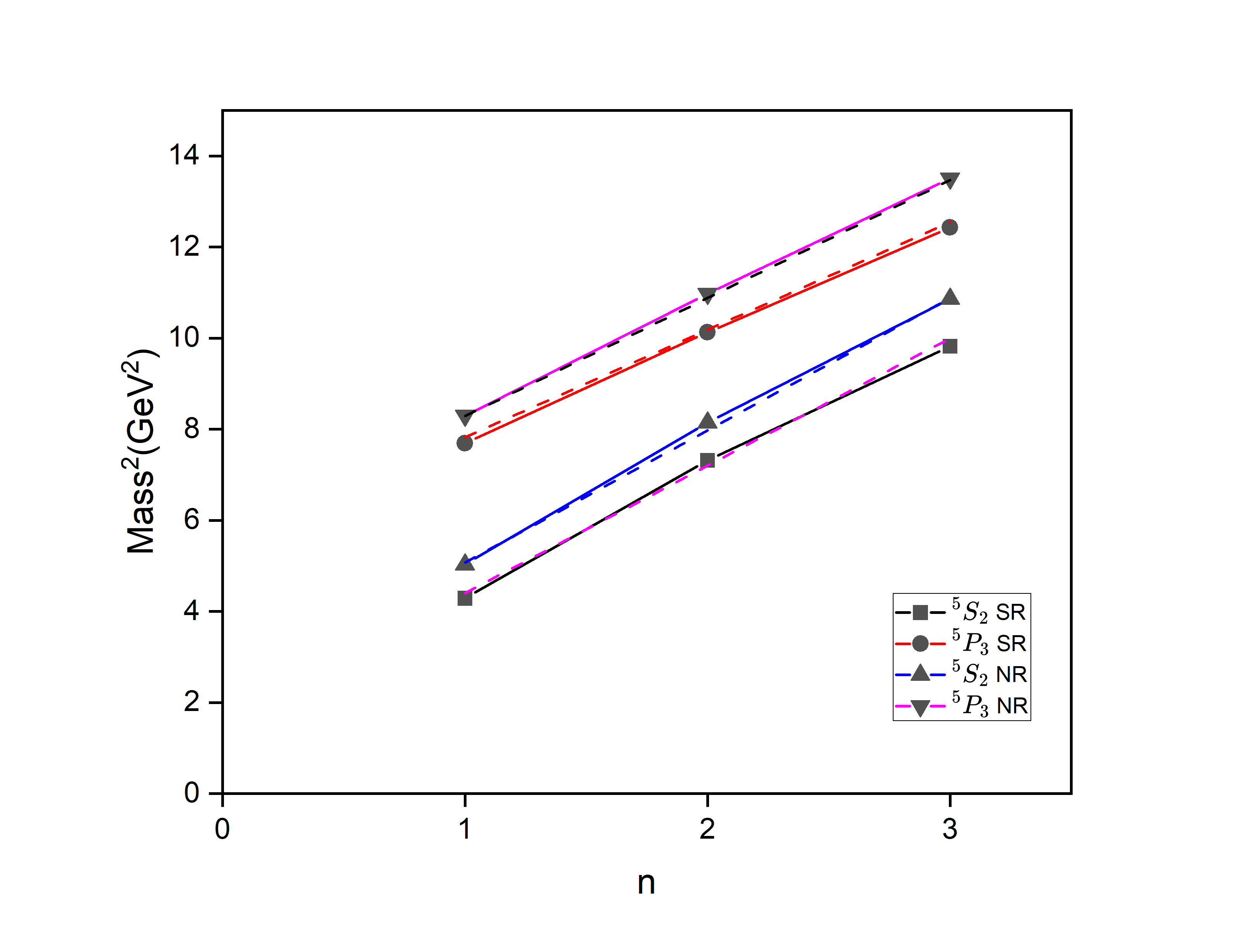}
		\caption{$ss\bar{q}\bar{q}$}
		\label{fig:tetraSSQQ2}
	\end{subfigure}
	\caption[]{Regge trajectory in the $(n, M^{2})$ plane for tetraquarks with Spin S = 2}
\end{figure*}

Understanding Regge trajectories is essential for analyzing the hadron spectrum and offers valuable insights into the strong interaction, particularly in the non-perturbative regime of Quantum Chromodynamics (QCD). They also play a key role in predicting new particles that fit into these trajectories. The Regge trajectories in the $(n,M^{2})$ plane are plots of the principal quantum number, n, against the square of the resonance mass, $M^{2}$. The Regge trajectories in the $(n, M^{2})$ plane are drawn for the Kaon with natural and
unnatural parity states, as shown in Figs. \ref{fig:mesonnrgraph1nat}, \ref{fig:mesonsrgraph1nat}, \ref{fig:mesonnrgraph2nat} and \ref{fig:mesonsrgraph2nat}. The solid straight lines represent our calculated results. Similarly, the trajectories in the $(n,M^{2})$ plane are depicted in Figs. \ref{fig:tetraSQSQS0}, \ref{fig:tetraSSQQ0}, \ref{fig:tetraSQSQS1}, \ref{fig:tetraSSQQ1}, \ref{fig:tetraSQSQS2} and \ref{fig:tetraSSQQ2} for $ss\bar{q}\bar{q}$ and $sq\bar{s}\bar{q}$ tetraquark with various spins.

\section{Results and Discussion}
\label{sec:5}
In the present work, the mass spectra of Kaon meson are calculated and tabulated in tables \ref{SWaveMesonmass} and \ref{GWaveMesonmass}. Using the fitting parameters obtained from the kaonic mass spectrum, masses of diquarks and anti-diquarks are calculated with various color configurations. Finally, the mass spectroscopy of all strange tetraquark in a semi-relativistic and non-relativistic framework using diquark-antidiquark formalism with various possible internal structures is calculated and presented in tables \ref{Swavetriplet} and \ref{Swavesextet} for antitriplet-triplet and sextet diquark-antidiquark, respectively. Table \ref{twomesonthreshold} shows various two-meson thresholds for different tetraquark states. In addition, the decay properties of tetraquarks in various decay channels have also been calculated. Using Fierz rearrangement, the spectator model, and heavy quarkonium annihilation, various decay channels for $T_{sq\bar{s}\bar{q}}$ and $T_{ss\bar{q}\bar{q}}$ tetraquarks are explored, which are shown in tables \ref{annihilationdecay} and \ref{spectatordecay}. The total decay rates for $T_{sq\bar{s}\bar{q}}$ and $T_{ss\bar{q}\bar{q}}$ strange tetraquarks are given in Table \ref{totaldecay}.

\subsection{Meson}

The calculated kaonic mass spectrum, when compared with various theoretical studies and with the PDG data, shows promising outcomes. For the S-wave, three resonances match the description of states calculated in the present work.
\subsubsection*{\textbf{A. $S-Wave$}}

The $1 ^{1}S_{0}$ and $3 ^{1}S_{0}$ states in the present study match the semi-relativistic and non-relativistic state descriptions. The $K^{\pm}$ are well established kaons with mass $493.77\pm0.016$ MeV, $I(J^{P})$ value $\frac{1}{2}(0^{-})$ and mean life $1.2380*10^{-8}$ s \cite{ParticleDataGroup:2022pth}. The $1 ^{1}S_{0}$ state of the present study has $I(J^{P})$ value $\frac{1}{2}(0^{-})$ and has a mass very close to $K^{\pm}$, 497 MeV for non-relativistic formalism and 496 for semi-relativistic formalism. The compared studies show $K^{\pm}$ mass very close to the PDG value, except ref \cite{Godfrey:1985xj} and \cite{Ebert:2009ub}, which predict a mass of 30 MeV and 15 MeV lower. The $K(1830)$ resonance also has $I(J^{P})$ value $\frac{1}{2}(0^{-})$ and is a fair candidate for the $3 ^{1}S_{0}$ state of present in semi-relativistic formalism. The $K(1830)$ resonance has been seen in $B^{+}\rightarrow J/\psi\phi K^{+}$ and $18.5K^{-}p\rightarrow3Kp. $ process with a decay width of $168\pm90^{+280}_{-104}$ MeV and $250$ MeV, respectively \cite{LHCb:2016axx,Bari-Birmingham-CERN-Milan-Paris-Pavia:1982xtk}. This resonance has an average mass of $1874\pm43^{+59}_{-119}$. The masses of radially excited scalar S-wave kaons show good agreement with compared studies except ref \cite{Ishida:1986vn}, which predicts lower masses.

The vector kaon $K^{*}(892)$ is also a very well established state with $I(J^{P})$ value  $\frac{1}{2}(1^{-})$ \cite{ParticleDataGroup:2022pth}. The average mass for $K^{*}(892)$ resonance observed in numerous hadroproduced processes including $0.9\bar{p}p\rightarrow K^{+}K^{-}\pi^{0}$, $8.25K^{-}p\rightarrow\bar{K^{0}}\pi^{-}p$, $200\pi p\rightarrow2K_{s}^{0}X$ and $70 K^{+}p\rightarrow K^{0}\pi^{+}X$ with an average decay width of $51.4\pm0.8$ MeV is $891.67\pm0.26$ MeV \cite{ParticleDataGroup:2022pth,CrystalBarrel:2019zqh,Birmingham-CERN-Glasgow-MichiganState-Paris:1984ppi,Napier:1984fg,Brussels-Genoa-Mons-Nijmegen-Serpukhov-CERN:1982pxs}. The average mass and average decay width for $K^{*}(892)$ resonance observed in numerous $\tau$ lepton decay processes, including $\tau^{-}\rightarrow K_{s}\pi^{-}\nu_{\tau}$ process, are $895.5\pm0.8$ MeV and $46.2\pm1.3$ \cite{Belle:2007goc}. The $1 ^{3}S_{1}$ state of the present study has mass 891 MeV and 899 MeV in non-relativistic and semi-relativistic formalism, respectively, with the $I(J^{P})$ value $\frac{1}{2}(1^{-})$ and fits the description of $K^{*}(892)$ resonance very well. The mass prediction for radially excited states for the S-wave vector kaon shows similar trends with the compared theoretical studies.
\subsubsection*{\textbf{B. $P-Wave$}}

The $K_{1}(1400)$ meson resonance has been observed in $\tau^{-}\rightarrow K^{-}\pi^{+}\pi^{-}\nu_{\tau}$, $11K^{-}p\rightarrow\bar{K^{0}}\pi^{+}\pi^{-}n$, $8.25K^{-}p\rightarrow{K^{0}_{s}}\pi^{+}\pi^{-}n$, $63K^{-}p\rightarrow K^{-}2\pi p$, $6K^{-}p\rightarrow\bar{K^{0}}\pi^{+}\pi^{-}n$ and $13K^{\pm}p\rightarrow(K\pi\pi)^{+-}p$ process with an average mass of $1403\pm7$ MeV and an average decay width of $174\pm13$ MeV. \cite {CLEO:2000nrp,Aston:1986jb,BIRMINGHAM-CERN-GLASGOW-MICHIGANSTATE-PARIS:1982skm,ACCMOR:1981yww,Etkin:1980me,Carnegie:1976cs}. This resonance has the $I(J^{P})$ value $\frac{1}{2}(1^{+})$ and seems a resonable fit for the $1 ^{1}P_{1}$ state of the present study in both formalisms. The mass spectrum of radially excited states of $^{1}P_{1}$ state in the present study shows higher mass when compared with other theoretical studies except ref \cite{Ishida:1986vn}. The mass spectrum of radial excitation of the $^{1}P_{1}$ state in ref \cite{Ishida:1986vn} shows good agreement with the non-relativistic formalism of the present study.

The $K(1630)$ resonance has an undefined $I(J^{P})$ value ($\frac{1}{2}(?^{?})$) and has been seen in $16.0\pi^{-}p\rightarrow(K^{0}_{s}\pi^{+}\pi)X^{+}\pi^{-}X^{0}$ with a decay width of $16^{+19}_{-16}$ MeV and a mass of $1629\pm7$ MeV \cite{Karnaukhov:1998qq}. The mass description of this state matches the $2^{3}P_{0}$ state of the current study in semi-relativist formalism very well. In the case of radial excitation of the $^{3}P_{0}$ state of the present study, the ground state and first excited state predict mass lower than the compared study, while the third and fourth states predict mass near that of other studies. The ground and excited state for the $^{3}P_{1}$ state predicts mass higher than the compared studies, except ref \cite{Ishida:1986vn}. The ground state of the $^{3}P_{1}$ state is lower in ref \cite{Ishida:1986vn}, but excited states show a comparable mass to the present study.

With an average mass of $1432.4 \pm1.3$ MeV for neutral resonance and $1427.3 \pm1.5$ MeV for charged resonance, the $K^{*}_{2}(1430)$ kaon resonance has $I(J^{P})$ value $\frac{1}{2}(2^{+})$. The neutral kaon for this resonance has been seen in $11K^{-}p\rightarrow K^{-}\pi^{+}n$, $11K^{-}p\rightarrow \bar{K^{0}}\pi^{+}\pi^{-}n$, $11K^{-}p\rightarrow \bar{K^{0}}2\pi n$, $13K^{\pm}p\rightarrow pK\pi$ and $8.25K^{-}p\rightarrow NK^{0}_{s}\pi\pi$ process with an average decay width of $109\pm5$ \cite{Aston:1987ir,Aston:1986jb,Aston:1984ns,BIRMINGHAM-CERN-GLASGOW-MICHIGANSTATE-PARIS:1982skm,Estabrooks:1977xe}. Similarly, the charged $K^{*}_{2}(1430)$ kaon resonance has been seen in $J/\psi\rightarrow K^{+}K^{-}\pi^{0}$, $8.25K^{-}p\rightarrow\bar{K^{0}}\pi^{-}p$, $30K^{+}p\rightarrow{K^{0}_{s}}\pi^{+}p$, $50K^{+}p\rightarrow{K^{0}_{s}}\pi^{+}p$, $6.5K^{-}p\rightarrow\bar{K^{0}}\pi^{-}p$, $10K^{\pm}p\rightarrow{K^{0}_{s}}\pi p$ and $K^{+}p\rightarrow{K^{0}}\pi^{+}p$ process with an average decay rate of $100.0\pm2.1$ MeV \cite{BESIII:2019apb,Birmingham-CERN-Glasgow-MichiganState-Paris:1984ppi,Cleland:1982td,Toaff:1981yk,Martin:1977ki,Barnham:1971jc}. This state shows a good resemblance to the $1 ^{3}P_{2}$ kaon state of the present study in semi-relativistic formalism. The $K_{2}^{*}(1980)$ kaon resonance also has $I(J^{P})$ value $\frac{1}{2}(2^{+})$. This resonance has an average mass of $1994^{+60}_{-50}$ MeV and has been seen in $\psi(2S)\rightarrow K^{+}K^{-}\eta$, $J/\psi\rightarrow K^{+}K^{-}\pi^{0}$ and $11K^{-}p\rightarrow\bar{K^{0}}\pi^{+}\pi^{-}n$ processes with an average decay width of $348^{+50}_{-30}$ MeV \cite{BESIII:2019dme,BESIII:2019apb,Aston:1986jb}. This kaonic resonance matches the description of the $2 ^{3}P_{2}$ and $1 ^{3}F_{2} $ state of the present study in both formalisms. The compared studies predict similar mass spectra for the ground and excited state of the $^{3}P_{2}$ state, except ref \cite{Godfrey:1985xj} and \cite{Ebert:2009ub} which give a lower prediction.

\subsubsection*{\textbf{C. $D-Wave$}}

The kaonic resonance $K_{2}(1770)$ is a well-established state with a $I(J^{P})$ value of $\frac{1}{2}(2^{-})$ and an average mass of 1773$\pm8$ MeV. This resonance has been observed in the $B^{+}\rightarrow J/\psi\phi K^{+}$ and $11K^{-}p\rightarrow K^{-}\omega p$ processes with an average decay width of $186\pm14$ MeV \cite{LHCb:2016axx,Aston:1993qc}. This resonance is an apt candidate for the $1 ^{1}D_{2}$ state in the non-relativistic formalism of the present study. All the compared studies, except ref \cite{Ebert:2009ub} estimate a mass slightly few MeVs higher for the $1 ^{1}D_{2}$ state. $K_{2}(2250)$ resonance also has a $I(J^{P})$ value $\frac{1}{2}(2^{-})$ but is not a well established particle. This state has an average mass of $2247\pm17$ MeV. been observed in the $18K^{-}p\rightarrow\Lambda\bar{p}X$, $8K^{-}p\rightarrow\Lambda\bar{p}X$ and $50K^{+}p\rightarrow\Lambda\bar{p}X$ processes with an average decay width of $180\pm30$ MeV \cite{Bari-Birmingham-CERN-Milan-Paris-Pavia:1983nwf,Birmingham-CERN-Glasgow-MichiganState-Paris:1980rkx,Cleland:1980ya}. This resonance is an excellent candidate for the $2 ^{1}D_{2}$ and $2 ^{3}D_{2}$ states in the non-relativistic formalism of the present study. 

The $K^{*}(1680)$ is also a well-established particle with a $I(J^{P})$ value of $\frac{1}{2}(1^{-})$. This particle has an average mass of $1718\pm18$ MeV and has been seen in $B^{+}\rightarrow J/\psi\phi K^{+}$, $11K^{-}p\rightarrow \bar{K}^{0}\pi^{+}\pi^{-} n$ and $11K^{-}p\rightarrow K^{-}\pi^{+} n$ processes with an average decay width of $322\pm110$ MeV \cite{LHCb:2016axx,Aston:1987ir,Aston:1986jb}. The $1 ^{3}D_{1}$ kaonic state in both formalisms of the present study matches the mass description of $K^{*}(1680)$. The $K(3100)$ resonance is a narrow peak observed in several ($\Lambda\bar{p}^{+}$ + pions) and ($\bar{\Lambda}{p}^{+}$+ pions) states in $\Sigma^{-}Be$ reactions by \cite{Bristol-Geneva-Heidelberg-Lausanne-Rutherford:1986mlv} and in $n$ and $npA$ reactions by \cite{EXCHARM:1992fpf}. This state has an average mass of 3100 MeVs and an undefined $I(J^{P})$ value ($?(?^{?})$). The $4^{3}D_{1}$ state in the non-relativistic formalism of the current study is a good candidate for this kaonic resonance.

The compared studies predict the ground state for all D-wave states near the prediction for the present study. While refs. \cite{Oudichhya:2023lva}  and \cite{Ebert:2009ub} predict the first radially excited D-wave state to be about 185 MeV lower, ref \cite{Ishida:1986vn} and \cite{Godfrey:1985xj} show a much closer estimate. Only ref \cite{Oudichhya:2023lva}  predicts the second and third radially excited states for D-wave kaon, which are at least a hundred MeV lower than the current study.

\subsubsection*{\textbf{D. $F-Wave$}}

The $K_{2}^{*}(1980)$ meson is a well-established particle and is a pretty good fit for the $1 ^{3}F_{2}$ state in both formalisms of the present study. The compared studies predict a higher mass value for all F-wave states by at least 100 MeVs. 

\subsubsection*{\textbf{F. $G-Wave$}}

As per PDG, the kaonic resonance $K_{5}^{*}(2380)$ is not a well-established particle and still needs confirmation. The $I(J^{P})$ value of this state is $1\frac{1}{2}$ and has an average mass of $2382\pm24$. This resonance has been observed in the $11K^{-}p\rightarrow K^{-}\pi^{+}n$ process with an average decay of $178\pm50$ MeV \cite{Aston:1986rm}. The $1 ^{3}G_{5}$ meson in the semi-relativistic formalism of the present study matches this description very well. The present study predicts that the masses of the G-wave kaonic state are lower than in the compared studies.

\subsection{Tetraquark}

In the present study, tetraquarks are explored as an extension of the meson concept, where a meson consists of a quark-antiquark pair. By analogy, the formalism used for calculating properties of mesons, such as mass spectra, annihilation decay rates, strong decay processes, strong decay constants, and Regge trajectories, is applied to the analysis of light-strange tetraquarks. Due to the absence of experimental data for light-strange tetraquarks from the Particle Data Group (PDG), the two-meson threshold serves as a reference point for comparison. Furthermore, since this work does not differentiate between up (\( u \)) and down (\( d \)) quarks, both isospin states \( I = 0 \) and \( I = 1 \) are taken into account in the analysis of various resonances. This approach ensures that the potential effects of isospin symmetry are not overlooked. The study also reveals that the tetraquark configuration \( sq\bar{s}\bar{q} \), where \( q \) represents a light quark, has a slightly higher mass compared to the \( ss\bar{q}\bar{q} \) configuration, with a difference ranging from 5 to 20 MeV in both formalism approaches. This mass difference could be indicative of the varying quark dynamics within the tetraquark system and may provide insight into the stability and decay patterns of these exotic states. Future experimental validation of these predictions will be crucial in enhancing our understanding of tetraquark structures and their role in hadron physics. In future work, we intend to integrate the isospin contribution into our examination of tetraquarks formed from light quarks.

\subsubsection*{\textbf{A. $S-Wave$}}

The $f_{0}(1770)$ resonance has a $I(J^{PC})$ value of $0(0^{++})$ and an average mass of $1784^{+16}_{-14}$ MeV. This resonance has been seen in $29\pi^{-}p\rightarrow n \omega\phi$, $J/\psi\rightarrow\gamma\omega\phi$ and $ \psi(2S)\rightarrow\gamma\pi^{+}\pi^{-}K^{+}K^{-}$ process with an average decay width of $161\pm21$ MeV \cite{Kholodenko:2020xgs,BESIII:2012rtd,BES:2005iaq}. $\pi\pi$, $K\bar{K}$, $\eta\eta$ and $\omega\phi$ decay modes of this resonance have been observed so far. This description matches very well with the $1 ^{1}S_{0}$ state of the $ss\bar{q}\bar{q}$ and $sq\bar{s}\bar{q}$ tetraquark in $\bar{\textbf{3}}-\textbf{3}$ configuration within the non-relativistic formalism of the present study. If this state is considered a candidate for $f_{0}(1770)$ resonance, the isospin of the $1 ^{1}S_{0}$ state can be predicted to be 0. The $f_{0}(1200-1600)$ resonance also has a similar $I(J^{PC})$ value but does not have a well-defined average mass. The K-matrix pole from combined analysis of $\pi^{-}p\rightarrow\pi^{0}\pi^{0}n$, $\pi^{-}p\rightarrow K\bar{K}n$ and $\bar{p}p\rightarrow\pi^{0}\pi^{0}\pi^{0},\pi^{0}\eta\eta,\pi^{0}\pi^{0}\eta$ at rest predicts the mass of this resonance to be $1530^{+90}_{-250}$ with a decay width of $560\pm40$ \cite{Anisovich:2002ij}. This description makes $f_{0}(1200-1600)$ an ideal candidate for $1 ^{1}S_{0}$ state of $sq\bar{s}\bar{q}$ and $ss\bar{q}\bar{q}$ tetraquark in $\bar{\textbf{3}}-\textbf{3}$ configuration within the semi-relativistic formalism. The X(1545) resonance has an undefined $I(J^{PC})$ value. This resonance was observed in $40\pi^{-}p\rightarrow K^{0}_{s}K^{0}_{s}n+m\pi^{0}$ with a decay width of $6.0\pm2.5$ MeV and a mass of $1545\pm3$ MeV \cite{Vladimirsky:2008zz}. This resonance is also an apt candidate for the $1 ^{1}S_{0}$ state of $sq\bar{s}\bar{q}$ and $ss\bar{q}\bar{q}$ tetraquark in $\bar{\textbf{3}}-\textbf{3}$ configuration within the semi-relativistic formalism. All of the calculated masses for the ground state of the two tetraquarks in both formalism and internal configuration for the $1 ^{1}S_{0}$ state are above the calculated two-meson threshold. Hence, $ss\bar{q}\bar{q}$ and $sq\bar{s}\bar{q}$ can decay strongly in two modes: $K_{0}^{\pm}+K_{0}^{\mp}$ and $\eta_{s}(1S)+\pi(1S)$. However, the sextet configuration for the $1 ^{1}S_{0}$ state in non-relativistic formalism does not cross the two-meson threshold for either mode. Ref. \cite{Kim:2023bac,Kim:2024adb} calculates the mass of tetraquark nonet, predicting two tetraquark with internal structure $(qs)\bar{q}\bar{s}$, namely $a_{0}(980)$ and $f_{0}(980)$. 

The $b_{1}(1960)$ and $h_{1}(1965)$ resonances have $I(J^{PC})$ values of $1(1^{+-})$ and $0(1^{+-})$ respectively. These resonances have been observed in a combined analysis of the meson channel with $I=0,1$ and $C=-1$ ($0.6-1.9p\bar{p}\rightarrow\omega\pi^{0}, \omega\eta\pi^{0},\pi^{+}\pi^{-}$) \cite{Anisovich:2002su}. The $b_{1}(1960)$ resonance has an average mass of $1960\pm35$ MeV and a decay width of $230\pm50$ MeV. The $h_{1}(1960)$ resonance has an average mass of $1965\pm45$ MeV and a decay width of $345\pm75$ MeV. The mass and $J^{PC}$ value of $ss\bar{q}\bar{q}$ and $sq\bar{s}\bar{q}$ tetraquark in non-relativistic formalism with the $\bar{3}-3$ internal structure of the present study fit very well with those of these two resonances. If $b_{1}(1960)$ resonance is considered as a $ss\bar{q}\bar{q}$ or $sq\bar{s}\bar{q}$ tetraquark candidate, it can be concluded that the isospin of the considered tetraquark is 1. Similarly, if the $h_{1}(1965)$ resonance is considered, the isospin can be taken as 0. The ground state of $^{3}S_{1}$ tetraquark has 3 decay modes for strong decay, which can be used as the two meson thresholds, namely $K_{1}^{*}(1S)K^{\pm}_{0}(1s)$, $\phi(1s)\pi(1S)$, and $\rho\eta_{s}$. All the calculated tetraquarks in the $1 ^{3}S_{1}$ state lie above these two meson thresholds.

The $f_{2}(2010)$ resonance has an average mass of $2011^{+62}_{-76}$ MeV and an average decay of $202\pm60$ MeV. This resonance is a well-established particle with a $I(J^{PC})$ value of $0(2^{++})$ and has been observed in the $22\pi^{-}p\rightarrow\phi\phi n$ process \cite{Etkin:1987rj}. The $1 ^{5}S_{2}$ tetraquark in the semi-relativistic formalism of the present study fits very well with this description; however, numerous studies predict this resonance as a strangeonium state. 

In the case of the $1 ^{5}S_{2}$ state of tetraquark in non-relativistic formalism, five resonances ($X(2210)$, $X_{2}(2210)$, $f_{J}(2220)$, $f_{2}(2240)$ and $a_{2}(2255)$) show great resemblance. The $I(J^{PC})$ values of $X_{2}(2210)$, $f_{2}(2240)$ and $a_{2}(2255)$ resonances are $0(2^{++})$, $0(2^{++})$ and $1(2^{++})$ respectively. The X(2210) resonance has an undefined $I(J^{PC})$ value, while the $f_{J}(2220)$ resonance has a half-defined $0(2^{++}\text{or}4^{++})$. The $f_{J}(2220)$ resonance has been seen in $e^{+}e^{-}\rightarrow J/\psi\rightarrow\gamma\pi^{+}\pi^{-}$, $e^{+}e^{-}\rightarrow J/\psi\rightarrow\gamma K^{+}K^{-}$, $e^{+}e^{-}\rightarrow J/\psi\rightarrow\gamma K^{0}_{s}K^{0}_{s}$, $e^{+}e^{-}\rightarrow J/\psi\rightarrow\gamma p\bar{p}$, $e^{+}e^{-}\rightarrow \gamma K^{+}K^{-}$, $e^{+}e^{-}\rightarrow \gamma K^{0}_{s}K^{0}_{s}$ and $11K^{-}p\rightarrow K^{+}K^{-}\Lambda$ process with an average decay width of $23^{+8}_{-7}$ MeV and an average mass of $2231.1\pm3.5$ MeV \cite{BES:1996upd,Aston:1988yp,MARK-III:1985qfw}. The $X(2210)$ resonance has been observed in the $10\pi^{-}p\rightarrow K^{+}K^{-}n$ and $11.2 \pi^{-}p$ processes with a mass of $2210^{+79}_{-21}$ MeV and $2207\pm22$ MeV, respectively \cite{BARI-BONN-CERN-DARESBURY-GLASGOW-LIVERPOOL-MILAN-VIENNA:1979msf,Caso:1970vv}. The decay widths of the $10\pi^{-}p\rightarrow K^{+}K^{-}n$ and $11.2 \pi^{-}p$ processes are estimated to be $203^{+437}_{-87}$ MeV and $130$ MeV, respectively.
$X_{2}(2210)$ has been observed in the fit of tensor partial waves from BES3 in the multipole basis, which might be a cluster of $J^{PC}=2^{++}$ resonances \cite{Klempt:2022qjf}. This resonance was observed in the $J/\psi\rightarrow\gamma\pi^{0}\pi^{0},\gamma K_{s}^{0}K_{s}^{0}$ process while searching tensor glueball. The $X_{2}(2210)$ has an average mass of $2210\pm60$ and a decay width of $360\pm120$ MeV. The $f_{2}(2240)$ resonance has been observed in the combined analysis of \cite{CrystalBarrel:1999zaz,Anisovich:1999pt,Anisovich:1999xm,Anisovich:1999ag,Anisovich:2000ae,Anisovich:2000ut} with an average mass of $2240\pm15$ and a decay width of $241\pm30$ MeV. The partial wave analysis of $\bar{p}p\rightarrow\eta\eta\pi^{0}$ done in the combined analysis of \cite{CrystalBarrel:1999zaz,CrystalBarrel:1999tdf,Anisovich:2001pn,Anisovich:2001pp} predicts the $a_{2}(2255)$ resonance, which has an average mass of $2255\pm20$ MeV and a decay width of $230\pm15$ MeV. The mass of the $1 ^{5}S_{2}$ state of $ss\bar{q}\bar{q}$ and $sq\bar{s}\bar{q}$ tetraquarks in non-relativistic formalism is 2243.27 and 2236.40 MeV, which makes its mass adjacent to all the resonances discussed so far. If the tetraquark is assigned isospin I = 1, then $a_{2}(2255)$ resonance is the appropriate choice. However, if the tetraquark is assigned a zero isospin, then $f_{J}(2220)$ and $f_{2}(2240)$ resonance are fair candidates. 

\subsubsection*{\textbf{B. $P-Wave$}}

The $\eta(2190)$ resonance has been observed in the broad $J^{P}=0^{-}$ meson in $J/\psi$ radiative decays with an average mass of $2190\pm50$ MeV and a decay width of $850\pm100$ MeV \cite{Bugg:1999jc}. This resonance is not a well-established particle but has a $I(J^{PC})$ value of $0({0^{-+}})$. This description matches the $1 ^{3}P_{0}$ state $sq\bar{s}\bar{q}$ and $ss\bar{q}\bar{q}$ tetraquark in semi-relativistic formalism, which have masses of 2147 MeV and 2133 MeV, respectively. 

The $\eta(2320)$ resonance was seen by the combined analysis of $\bar{p}p\rightarrow\eta\eta\eta$ from \cite{Anisovich:2000ix} and $\bar{p}p\rightarrow\eta\pi^{0}\pi^{0}$ from \cite{Anisovich:2000ut} with a mass of $2320\pm15$ MeV and a decay width of $230\pm35$ MeV. The $I(J^{PC})$ value of this resonance is $0(0^{-+})$. The masses of $1 ^{3}P_{0}$ state $sq\bar{s}\bar{q}$ and $ss\bar{q}\bar{q}$ tetraquark in non-relativistic formalism have masses of 2305 MeV and 2315 MeV respectively, which make them an ideal candidate for this resonance. The $\pi(2360)$ resonance has been enhanced by the partial wave analysis of $p\bar{p}$ annihilation channels in flight with $I=1$ and $C=+1$ with a decay width of $300^{+100}_{-50}$ in $2.0\bar{p}p\rightarrow3\pi^{0},\pi^{0}\eta,\pi^{0}\eta^{'}$ processes \cite{Anisovich:2001pn}. The mass and $I(J^{PC})$ value of this resonance are $2360\pm25$ and $1(0^{-+})$, respectively. These values also match the $1 ^{3}P_{0}$ state $sq\bar{s}\bar{q}$ and $ss\bar{q}\bar{q}$ tetraquark in non-relativistic formalism.

The $2 ^{3}P_{0} $ state $sq\bar{s}\bar{q}$ and $ss\bar{q}\bar{q}$ tetraquark in semi-relativistic formalism with $\textbf{6}-\bar{\textbf{6}} $ The color structure has four resonances as potential candidates, namely, $\eta(2010)$, $\pi(2070)$, $\eta(2100)$ and $X(2100)$. The $\eta(2010)$ has an average mass of $2010^{+35}_{-60}$ MeV and a decay width of $270\pm60$ MeV. The $\pi(2070)$ was observed along with the $\pi(2360)$ resonance mentioned before with the same decay width and a mass of $2070\pm35$. The $\eta(2100)$ resonance has been observed in two different processes. When derived from a partial wave analysis of $J/\psi\rightarrow\gamma\phi\phi$, for which the primary signal is $\eta(2225)\rightarrow\phi\phi$, the $\eta(2100)$ resonance is observed in $J/\psi\rightarrow\gamma K^{+}K^{-}K^{+}K^{-}$ with mass $2050^{+30+75}_{-24-26}$ MeV and decay width $250^{+36+181}_{-30-164}$ MeV \cite{BESIII:2016qzq}. However, when seen in the $J/\psi\rightarrow4\pi\gamma$ process, the $\eta(2100)$ resonance has an average mass of $2103\pm50$ MeV and a decay width of $187\pm75$ MeV \cite{DM2:1988esg}. The $X(2100)$ resonance was seen in the $100\pi^{-}p\rightarrow2\eta X$ process with a mass $2100\pm40$ MeV and a decay width of $250\pm40$ MeV \cite{Serpukhov-Brussels-LosAlamos-AnnecyLAPP:1985qjh}. The $I(J^{PC})$ value for $\eta(2100)$ and $\eta(2010)$ is $0(0^{-+})$ and suggests the tetraquark has an isospin 0. On the other hand, $\pi(2070)$ has a $I(J^{PC})$ value of $1(0^{-+})$, suggesting the tetraquark's isospin zero. The $2 ^{3}P_{0}$ state $sq\bar{s}\bar{q}$ and $ss\bar{q}\bar{q}$ tetraquark in semi-relativistic formalism with $\textbf{6}-\bar{\textbf{6}}$ color structure in the present study has masses of 2084 MeV and 2070 MeV, respectively. The four resonances match the description of $2 ^{3}P_{0}$ states very well, making them potential candidates. 

The $1 ^{5}P_{1}$ state $sq\bar{s}\bar{q}$ and $ss\bar{q}\bar{q}$ tetraquark in non-relativistic formalism suffer the same fate as the $2 ^{3}P_{0}$ state and the $1 ^{5}S_{2}$ state with multiple resonances as potential candidates. The $1 ^{5}P_{1}$ has a mass of 2304 MeV and 2314 MeV for $sq\bar{s}\bar{q}$ and $ss\bar{q}\bar{q}$ tetraquark in non-relativistiv formalism, respectively, with a $J^{PC}$ value of $1^{--}$. The $\rho(2270)$ resonance has been seen in $0.6-1.9\bar{p}p\rightarrow\omega\pi^{0},\omega\eta\pi^{0},\pi^{+}\pi^{-}$ with a mass of $2265\pm40$ MeV and decay width of $325\pm80$ MeV \cite{Anisovich:2002su}. When observed in the $20-70 \gamma p\rightarrow p\omega\pi^{+}\pi^{-}\pi^{0}$ process, the $\rho(2265)$ resonance has a mass prediction of $2280\pm50$ and a decay width of $440\pm110$ MeV \cite{OmegaPhoton:1985vyt}. The partial wave analysis of the data on $p\bar{p}\rightarrow\bar{\Lambda}\Lambda$ from \cite{Barnes:2000be} predicts the mass and decay width of $\omega(2290)$ resonance to be $2290\pm20$ MeV and $275\pm35$ respectively \cite{Bugg:2004rj}. Similarly, the photon diffractive dissociation to $\rho\rho\pi$ and $\rho\pi\pi\pi$ states in $25-50\gamma p\rightarrow\rho^{\pm}\rho^{0}\pi^{\mp}$ predicts the $\omega(2330)$ resonance with a mass of $2330\pm30$ MeV and a decay width of $435\pm75$ MeV \cite{OmegaPhoton:1988guj}. Considering the calculated masses and $J^{PC}$ of $1 ^{5}P_{1}$ state $sq\bar{s}\bar{q}$ and $ss\bar{q}\bar{q}$ tetraquark, the resonances discussed here contest as potential candidates. If the resonances $\omega(2290)$ and $\omega(2230)$ are considered, the tetraquark must have isospin zero since their $I(J^{PC})$ value is $0(1^{--})$. However, considering $\rho(2265)$ as a tetraquark candidate suggests the tetraquark with isospin 1, given that the $I(J^{PC})$ of $\rho(2265)$ is $1(1^{--})$.

The $1 ^{5}P_{1}$ state $sq\bar{s}\bar{q}$ and $ss\bar{q}\bar{q}$ tetraquark in semi-relativistic formalism for the present study has 2153 MeV and 2139 MeV, respectively. The $\rho(2150)$ resonance is a not well-established resonance that was earlier referred to as $T_{1}(2190)$ \cite{ParticleDataGroup:2022pth}. This resonance has been seen in numerous $e^{+}e^{-}$ processes \cite{BESIII:2020xmw,BESIII:2020kpr,BESIII:2019dme}, $p\bar{p}\rightarrow\pi\pi$ process \cite{Hasan:1994he,Oakden:1993he}, S-channel $N\bar{N}$ process \cite{Anisovich:2002su,Cutts:1974vi} and $\pi^{-}p\rightarrow\omega\pi^{0}n$ process \cite{GAMS:1994jop,IHEP-IISN-LANL-LAPP-KEK:1992puu}. No average mass is available in PDG but 2150 MeV can be taken as the central value for its mass. This state has a $I(J^{PC})$ value of $1(1^{--})$, which suggests that this resonance can be a good tetraquark candidate with isospin 1. Ref. \cite{Agaev:2020zad,Agaev:2019coa} speculates $Y(2175)$ as a light vector tetraquark state with a mass of $2173\pm85$ MeV and a decay width of $91.1\pm20.5$ MeV, with a strong emphasis on the internal structure of the resonance being $[su][\bar{s}\bar{u}]$. Similarly, ref.  \cite{Wang:2006gj} and \cite{Xin:2022qnv} predict an explanation of the P-wave mass of light strange vector tetraquark state between $2.06\pm0.13$ GeV and $3.52\pm0.11$ GeV for the P-wave with partial derivative and $2.06\pm0.13$ GeV to $3.54\pm0.09$ GeV for the P-wave with covariant derivative. Ref. \cite{Xin:2022qnv} predicts $Y(2175),X(2240)$ and $X(2400)$ as tetraquark candidates.  

$X(2600)$ resonance has undefined $I(J^{PC})$ value and is seen in $J/\psi\rightarrow\gamma\pi^{+}\pi^{-}\eta^{'}$ process with an average mass and decay width of $2618.3\pm2.0^{-16.3}_{-1.4}$ MeV and $195\pm5^{+26}_{-17}$ MeV \cite{BESIIICollaboration:2022kwh}. This resonance has been observed to decay in $f_{0}(1500)\eta^{'}$ mode and $X(1540)\eta^{'}$ mode, making a very possible candidate for $ss\bar{q}\bar{q}$ and $sq\bar{s}\bar{q}$ tetraquark. Since it has an undefined $J^{PC}$ value, it becomes very difficult to point out the exact state of this resonance. However, it is safe to predict that this resonance is a P-wave state.
\section{Conclusion} 

To summarize, utilizing fitting parameters determined from Kaon mass spectra in a semi-relativistic and non-relativistic framework, we have calculated the mass spectra of $sq\bar{s}\bar{q}$ and $ss\bar{q}\bar{q}$ tetraquarks in different diquark-antidiquark configurations and internal structures, including relativistic corrections. The computed states have corresponding $J^{PC}$ values assigned to them. Furthermore, we have used the annihilation model \cite{Kher:2018wtv}, spectator model \cite{Becchi:2020mjz,Becchi:2020uvq}, and rearrangement model \cite{Ali:2019roi} to analyze the decay processes of Kaons and $sq\bar{s}\bar{q}$ and $ss\bar{q}\bar{q}$ tetraquarks. Additionally, we have determined two-meson thresholds in both formalisms for tetraquarks. For various spins and parity states, regge plots for Kaon states and strange tetraquarks have been plotted.

Various resonances have been investigated as potential candidates for S-wave Kaon states, including $K(497)$, $K(1830)$, and $K^{*}_{1}(891)$. For P-wave Kaon states, we examined $K_{1}(1400)$, $K(1630)$, $K_{2}^{*}(1430)$, and $K_{2}^{*}(1980)$. D-wave candidates like $K_{2}(1770)$, $K_{2}(2250)$, $K^{*}(1680)$, $K(3100)$, and $K_{2}(2250)$ were explored, along with $K^{*}_{2}(1980)$ as an F-wave candidate and $K_{5}^{*}(2380)$ as a G-wave candidate. For S-wave tetraquark states $ss\bar{q}\bar{q}$ and $sq\bar{s}\bar{q}$, potential candidates include $f_{0}(1770)$, $X(1545)$, $b_{1}(1960)$, $h_{1}(1965)$, $f_{2}(2010)$, $X(2210)$, $X_{2}(2210)$, $f_{J}(2220)$, $f_{2}(2240)$, and $a_{2}(2255)$. Similarly, potential P-wave candidates such as $\eta(2190)$, $\eta(2320)$, $\pi(2360)$, $\eta(2100)$, $\eta(2010)$, $\pi(2070)$, $X(2100)$, $\rho(2270)$, $\rho(2265)$, $\omega(2290)$, $\omega(2330)$, $\rho(2150)$, and $X(2600)$ resonances were explored.

These findings could potentially serve as valuable input and comparative data for institutions such as PANDA, J-PARC, Belle, LHCb, and others that focus on in-depth analyses of resonances involving strange quarks. This investigation provides a robust foundation for future studies employing the formalism described here to investigate tetraquarks with various quark flavors. Additionally, this research paves the way for more comprehensive theoretical models and experimental validations, thereby enhancing our understanding of hadronic structures and interactions. The methodologies and results presented could also contribute to the development of new experimental techniques and the refinement of existing ones, facilitating more precise measurements and discoveries in the field of particle physics.

\section{Data Availability Statement} The datasets generated during and/or analysed during the current study are available from the corresponding author on reasonable request.

\end{document}